




%
\input harvmac.tex






\font\tenmsy=msbm10
\font\sevenmsy=msbm10 at 7pt
\font\fivemsy=msbm10 at 5pt
\newfam\msyfam 
\textfont\msyfam=\tenmsy
\scriptfont\msyfam=\sevenmsy
\scriptscriptfont\msyfam=\fivemsy
\def\blackB{\fam\msyfam\tenmsy}
\def\ZZ{{\blackB Z}}

\def\RR{{\blackB R}}

\def\NN{{\blackB N}}




\def\frac#1#2{{\textstyle{#1\over #2}}}

\def\b#1{\kern-0.25pt\vbox{\hrule height
0.2pt\hbox{\vrule
width 0.2pt \kern2pt\vbox{\kern2pt
\hbox{#1}\kern2pt}\kern2pt\vrule
width 0.2pt}\hrule height 0.2pt}}
\def\ST#1{\matrix{\vbox{#1}}}
\def\STrow#1{\hbox{#1}\kern-1.35pt}
\def\bv{\b{\phantom{1}}}





\def\eqalignD#1{
\vcenter{\openup1\jot\halign{
\hfil$\displaystyle{##}$~&
$\displaystyle{##}$\hfil~&
$\displaystyle{##}$\hfil\cr
#1}}
}
\def\eqalignT#1{
\vcenter{\openup1\jot\halign{
\hfil$\displaystyle{##}$~&
$\displaystyle{##}$\hfil~&
$\displaystyle{##}$\hfil~&
$\displaystyle{##}$\hfil\cr
#1}}
}

\def\text#1{\quad\hbox{#1}\quad}
\def\gh{\hat{g}}

\def\la{\lambda}

\def\e{\epsilon}
\def\nuh{{\hat \nu}}
\def\muh{{\hat \mu}}

\def\E{\widehat {E}}
\def\Ec{{\cal \E}}

\def\A{\widehat {A}}

\def\B{\widehat {B}}
\def\C{\widehat {C}}
\def\D{\widehat {D}}

\def\F{\widehat {F}}

\def\lah{{\hat \lambda}}

\def\gh{{\widehat g}}
\def\rw{\rightarrow}

\def\lra{\leftrightarrow}
\def\su{\widehat{su}}

\def\sp{\widehat{sp}}

\def\Nc{{\cal N}}

\overfullrule=0pt

\newcount\eqnum
\eqnum=0
\def\eq{\eqno(\secsym\the\meqno)\global\advance\meqno
by1}
\def\eqlabel#1{{\xdef#1{\secsym\the\meqno}}\eq }

\newwrite\refs
\def\startreferences{
     \immediate\openout\refs=references
     \immediate\write\refs{\baselineskip=14pt \parindent=16pt \parskip=2pt}
}
\startreferences

\refno=0
\def\aref#1{\global\advance\refno by1
     \immediate\write\refs{\noexpand\item{\the\refno.}#1\hfil\par}}
\def\ref#1{\aref{#1}\the\refno}
\def\refname#1{\xdef#1{\the\refno}}

\newcount\exno
\exno=0
\def\Ex{\global\advance\exno by1{\noindent\sl Example \the\exno:

\nobreak\par\nobreak}}

\parskip=6pt

\font\sm cmr6



\Title{\vbox{\baselineskip12pt
\hbox{
}}}
{\vbox {\centerline{Fusion bases as facets of
polytopes   }}}
\centerline{L. B\'egin$^*$,
     C. Cummins$^\sharp$
L. Lapointe$^\dagger$  and P. Mathieu$^\natural$
}

\smallskip\centerline{* \it Secteur Sciences, Campus
d'Edmundston, Universit\'e de
Moncton, N.-B., Canada, E3V 2S8}
\smallskip\centerline{$^\sharp$ \it Mathematics
Department, University of
Concordia, Montr\'eal, Canada H3G 1M8}
\smallskip\centerline{$^\dagger$ \it Department of Mathematics and
Statistics, McGill University,
Montr\'eal, Qu\'ebec, Canada H3A 2K6}\smallskip\centerline{$^\natural$ \it
D\'epartement de Physique,
Universit\'e Laval, Qu\'ebec, Canada G1K 7P4}
\noindent

\bigskip

{\bf Abstract}: A new way of constructing
fusion bases (i.e., the set of inequalities governing
     fusion rules) out
of fusion elementary couplings is presented.  It relies
on a polytope reinterpretation
of the problem: the elementary couplings are
associated to the vertices of the
polytope while the inequalities defining the fusion
basis are the facets. The
symmetry group of the polytope associated to the
lowest rank affine Lie
algebras is found; it has order 24 for
$\su(2)$, 432 for
$\su(3)$ and quite surprisingly, it reduces to 36
for $\su(4)$, while it
is only of order 4 for $\sp(4)$.  This drastic reduction in
the order of the symmetry group
as the algebra gets more complicated is rooted in the
presence of many linear
relations between the elementary couplings that break
most of the potential
symmetries. For $\su(2)$ and $\su(3)$, it is shown that the
fusion-basis defining
inequalities can  be generated from few (1 and 2
respectively) elementary ones. For $\su(3)$, new symmetries
of the fusion coefficients are found.

     \Date{05/00-08/01\ \ (hepth@xxx/0108213)}



\newsec{Introduction}


Affine fusion rules give the number of integrable
representations $\nuh$ that appear in the product of
two integrable
representations $\lah$ and $\muh$ for a given affine
algebra $\gh$ at fixed
level
$k$ (see e.g., [\ref{P. Di Francesco, P.
Mathieu and D.
S\'en\'echal, {\it
Conformal field theory}, Springer 1997.}\refname\CFT] Chapter 16).
Fusions are in fact truncated finite Lie algebra tensor
products, with the degree of
truncation fixed solely by the level. More precisely,
fusion rules are
completely
characterized by the  tensor-product coefficients
pertaining to the
corresponding finite (i.e., non-affine)
representations and the set of
threshold
levels  [\ref{C.J. Cummins, P. Mathieu and M.A.
Walton, Phys. Lett. {\bf B254}
(1991) 390.}\refname\CMW].  The threshold level of a
particular  coupling
representing  one of the various copies of the
representation $\nuh$ in the
product
$\lah\times\muh$ is the lowest level at which this
coupling appears in this
fusion. (Note that only the full set of threshold levels
associated to a given triple
coupling is  an observable; the association of a
particular threshold level with
a given coupling is basis dependent.)

Even for $\su(N)$, no genuine combinatorial
methods -- analogous to the Littlewood-Richardson
rule --  have been found.  The
closest approach to such a goal has been obtained in
[\ref{L. B\'egin, C.
Cummins
and P. Mathieu, {\it Generating-function method for
fusion rules},
math-ph/0005002, J. Math. Phys. {\bf 41} 7640
(2000).}\refname\BCM] in which
     a new approach to the problem of fusion
rules was introduced,
centered on the notion of {\it fusion basis}. A
fusion basis is simply a
complete
set of inequalities, formulated in terms of a
complete set of variables
     needed to describe a tensor product, augmented with
an extra
variable, the level $k$. Examples of bases have been
constructed for
$\su(2)$, $\su(3)$, $\su(4)$
and $\sp(4)$.

The idea of the construction in [\BCM] is the
following: one
starts from the tensor-product elementary couplings,
extends this set to a
complete set of fusion elementary couplings (using,
for instance, the
conjectural
completeness under outer automorphism -- but there
exist other
possibilities) and
from these, construct the inequalities in terms of
the basis variables for which
the elementary couplings  are the elementary
solutions.

In [\BCM], the transition from the elementary
couplings to the inequalities uses
      Farkas' lemma.  The aim of this note is to indicate
another
way of reconstructing the fusion basis given the
fusion elementary couplings.
This new construction relies on a reinterpretation of
the fusion-rule
computations in terms of counting
points inside a polytope.
  A
polytope can be described by its vertices or its
facets.  The
reconstruction of the facets of a polytope from  its
vertices is the essential
trick we want to adapt to the problem of fusion
rules.  In our context, the
vertices are represented by the fusion elementary
couplings and the facets are
the inequalities for which the elementary couplings
are the elementary
solutions.
This reformulation is not purely cosmetic:
it allows us to use  powerful (e.g., computer)
techniques developed for
the study
of polytopes, for instance for deriving the facets
from its vertices.

But this
conceptual shift in the description of the fusion
basis has an immediate
benefit:
having constructed a polytope it is natural to look for
its symmetries.
This means
looking for the symmetry group of the fusion basis
and organising the
various inequalities into a number of orbits of the
group.
In this paper we find the
symmetry
group of the polytopes associated to the known fusion
bases.

\newsec{The $\su(2)$ example}


As a simple
illustrative
example we present the $\su(2)$ fusion basis:
$$ k\geq \la_1+n_{11}\qquad n_{12}\geq 0\qquad
\la_1\geq n_{12}\qquad
n_{11}\geq 0.\eqlabel\inedeux$$
The last three conditions define the Littlewood-
Richardson
(LR) basis, which is thus recovered from the fusion
basis in the infinite level
limit ($k\rightarrow \infty$).
This basis
describes
the solution set of the fusion $\lah\times\muh$ at
fixed positive integer
level $k$.
The two Dynkin labels of $\lah$ are $\la_0= k-\la_1$
and $\la_1$, with $\la_1$ being the
finite Dynkin label (and we will often use the
Dynkin label
notation: $\lah= [\la_0,\la_1]$). Both Dynkin labels
are assumed to be
non-negative integers.  The LR algorithm starts by
filling the boxes of the
first row
of the Young tableau associated to
$\mu$ with 1's, the second row with 2's, etc. For
$su(2)$, $\mu$ has only
one row,
containing
$\mu_1$ boxes. These boxes are inserted into the
tableau representing
$\la$, which
is a row of
$\la_1$ boxes: $n_{11}$ boxes are then added in the
first row and $n_{12}$ boxes
in the second row.  Therefore $n_{11}$ and $n_{12}$
are non-negative
integers and $n_{11}+n_{12}=
\mu_1$. Moreover, columns with two 1's are not
permitted, which forces
$\la_1\geq
n_{12}$. Finally, the tableau  associated to the
representation
$\nu$ is read off the resulting LR tableau by taking
out the columns of two
boxes:
$\nu_1=
\la_1+n_{11}-n_{12}$. (For more details on the LR
algorithm, see e.g.,
[\ref{I.G.MacDonald, {\it Symmetric Functions and
Hall Polynomials}, 2nd Edition, Oxford University Press
1995.}\refname\Mac,
\ref{M.Couture, C.J.Cummins and R.T.Sharp, J.Phys
{\bf A23} (1990)
1929.}\refname\CCS, \ref{L. B\'egin, C. Cummins and
P. Mathieu, {\it
Generating-function method for tensor products}, math-
ph/0005003, J.
Math. Phys. {\bf 41} 7640 (2000) (or hep-th/9811113
for an extended
version).}\refname\BCMa].)

The elementary solutions of this system of
inequalities, written in
terms of  vectors with entries ordered as $(k,\la_1,
n_{11}, n_{12})$, are
$$\eqalignT{  & \e_0= (1,0,0,0) \qquad &\qquad\E_0 :
d &\cr
     & \e_1= (1,1,0,1)&\qquad \E_1: dL_1N_{12}:&\quad
\ST{\STrow{\bv}\STrow{\b1}} \cr
     & \e_2= (1,1,0,0) &\qquad \E_2: dL_1 :&\quad
\ST{\STrow{\bv}}\cr
      & \e_3= (1,0,1,0)&\qquad\E_3: dN_{11}:&\quad
\ST{\STrow{\b1}}\cr}
\eqlabel\eleuu$$
We have also presented the corresponding LR tableau
at the right
and its `exponential' description in between (where
the variables $k,\,\la_1,\,
n_{11},\, n_{12}$ appear respectively as the
exponents of the dummy variables
$d,\,L_1,\, N_{11},\, N_{12}$). The problem we
consider here is the following:
given $\E_i$, $i=0,1,2,3$, how can we
reconstruct the inequalities? In
other words,  how to go from the vertices (\eleuu) to
the facets (\inedeux)?

\newsec{The polytope interpretation of Farkas' lemma}

As we just mentioned, for the present work we suppose
that the complete set of
fusion elementary couplings $\{\Ec_i\}$ is known.
These are expressed in
terms of
a set of variables, denoted collectively as $X_j$,
$j=1,\cdots,n$ (which are the
dummy variables
$d,\,L_1,\, N_{11},\, N_{12}$ in the previous
example), that furnish a
complete
description of the fusion rules.  These are in fact
the variables that describe
the tensor products with the addition of the extra
variable $d$ that keeps track
of the level of the affine algebra under
consideration.

A general coupling can always be decomposed into a
product of elementary couplings (and that this
decomposition may not be
unique is
irrelevant at this point).  This {product}
decomposition can be transferred
into a sum decomposition by characterising an
elementary coupling by the
exponents
$\e_{ij}$ in its decomposition in terms of the
variables $X_j$: an elementary
coupling is thereby associated to a $n$-dimensional
vector $\e_i$.

      Again, our problem  is the following: what is the
set
of linear and homogeneous inequalities for which the
$\e_i$ are the elementary
solutions? These inequalities will be formulated in
terms of the variables
$x_j$,
whose exponential versions are the $X_j$. The set
$\{x_i\}$ will typically
contain the finite Dynkin labels of the three affine
weights entering in the
fusions, together with the missing labels appropriate
for a complete description
of the corresponding tensor product, plus the level
$k$.

Any coupling $\prod_i X_i^{\,x_i}$ can thus be
decomposed  in the form
$\prod_i \Ec_i^{\,a_i}$.
With $\Ec_i =\prod_j X_j^{\e_{ij}}$, reading off a
particular coupling means looking for a choice set of
non-negative integers $\{a_i\}$ fixed by
$$\sum_i a_i\e_{ij}  = x_j\eq$$
in terms of non-negative integers $x_j$.
    We are thus
searching
for the conditions ensuring the existence
of such  a coupling. This is precisely what Farkas'
lemma can do.
Quite remarkably, the lemma gives  existence
conditions in the form of
inequalities  on the $x_j$'s and these are precisely
the inequalities we
are looking
for.

If we write $V_{ij}=\e_{ji}$ (hence, the columns of
the matrix $V$ are the
elementary solutions), Farkas' lemma gives existence
conditions for solutions of
$$ V\, a= x\eqlabel\basic$$
(in matrix form). For instance, for $\su(2)$, $V$
takes the form (cf.
(\eleuu)):
$$V= \pmatrix {1&1&1&1\cr 0&1&1&0\cr 0&0&0&1\cr
0&1&0&0\cr}\eqlabel\vdeu$$
     In fact, we are interested in the integer
solutions of (\basic).

We should stress that Farkas' lemma is true in
general only in the rational
case.  Sufficient conditions for its application to
the integer case
are known (see for instance  [\ref{A.
Schrijver, {\it Theory of linear and integer
programming}, Wiley
1986.}\refname\Schri]), but these are not satisfied by
our matrices $V$.
Hence, the analysis has to be completed by a
verification step, that is, to
check
that the elementary solutions of the inequalities
obtained are the elementary
solutions used at the start (cf. the discussion in
sect. 3 of [\BCM]).

The polytope interpretation of (\basic) is almost
immediate (see the next
section): modulo renormalization, (\basic) is the
equation of a
polytope whose vertices $v_i$ are the
columns of the matrix $V$ without its first row,
hence our elementary couplings
without their first entry $d$.

Now it is a well-known result (the Weyl-Minkowski
theorem)
that a polytope
can be described either in terms of its vertices or
its facets.
In the present case, we have the vertices;  the
question is thus: what are the
facets of the polytopes whose vertices are our fusion
elementary couplings?
These
facets are the sought for inequalities, which form
the fusion
basis.

This reformulation is  made more
formal in the next section. The following sections
are devoted to the study of
the symmetry group of the fusion polytopes for the
simplest affine Lie algebras.

\newsec{Formalizing the polytope reinterpretation}

To recast the problem in a general setting, let $V
\in {\NN}^{n\times m}$
be such that
$$
V \, =\, \pmatrix{v_1 \dots v_{m}\cr} \, =
\pmatrix {w_1 & \cdots & w_m \cr {\bf{v}}_1 & \cdots
& {\bf{v}}_m \cr}
\, , \eq
$$
with $v_1,\dots,v_m$ the fusion elementary couplings
and $w_1,\dots,w_m$ the
entries in the first row of $V$.  Also, let $\cal S$
denote the set of fusion
couplings
$$
{\cal S} =\Bigl\{ x = Va=a_1 v_1 +\cdots+a_m v_m \,
\Big| \, a_1,\dots,a_m
\in \NN
\Bigr\}\, ,
$$
which, from our central hypothesis, is the set of
nonnegative integral
combinations of elementary
couplings.   It is natural to suppose that the fusion
couplings
are the integral points of a certain geometric
object $E$, that is,
the vectors $v_1,\dots,v_m$ form a Hilbert basis
of $E$. (A finite set of vectors $v_1,\dots,v_m$ is a Hilbert basis
of $C$=cone($a_1,\dots,a_t$) if $C\cap \ZZ^n = \NN v_1 + \dots +
\NN v_m$.)
One obvious choice for $E$, and the one we will make,
is the cone generated by
$v_1,\dots,v_m$, that is
$$
E= \left\{ x=V \lambda = \lambda_1 v_1 +\cdots
+\lambda_m v_m \, \Big| \,
\lambda_1,\dots,
\lambda_m \in {\RR}_+  \right\} \, .
\eqlabel\cone
$$
In general, $v_1,\dots v_m$ may or may not be a Hilbert basis
of this cone. However, for our examples
we find  that
$v_1,\dots,v_m$ is
indeed   a Hilbert basis of $E$.
This checking process is equivalent to the
verification step mentioned in
the last section.
The set ${\cal S}$ is thus given by
$$
{\cal S} = E \cap \NN^n \, ,
\eq
$$
%
%
and the fusion inequalities are simply the facets of
the cone $E$.  We can therefore use Farkas' lemma (or
any other method)
to obtain the facets of $E$.
Now, it turns out that
the fusion inequalities
are also the
facets of a polytope.  The remainder
of this section is devoted to this reinterpretation.

If we write the vectors $x \in E$ as
$$
x = \pmatrix {x_0 \cr {\bf{x}} \cr} \, ,
\eq
$$
we have that the fusion couplings at level $k$ are
the integral points of the
space, $P^{(k)}$, corresponding to
the intersection
between the hyperplane $x_0=k$ and the cone $E$.  If
we drop the
$x_0$ component, which has value $k$ in $P^{(k)}$, we
can describe $P^{(k)}$
as
$$
\eqalign{
{{P}}^{(k)} & = \left\{ {\bf{x}} \, \Big|\, \pmatrix {k
\cr {\bf{x}}\cr} \in
E
\right\} \, = \, \left \{ {\bf{x}} \, \Big | \,
\pmatrix {k \cr
{\bf{x}}\cr} =V\,
\lambda \, , {{\lambda}} \in
{\RR}^{m}_+ \, \right \}  \, \cr & =  \,  \Big \{ {\bf{x}}=
\sum_i {\bf{v}}_i \,
{{\lambda_i}} \, \big | \, \lambda \in
     {\RR}^{m}_+ \,
, \, \sum_i \lambda_i w_i=k \, \Big \} \, .\cr}
\eq
$$
The integral points of $P^{(k)}$ are
$$
{\cal S}^{(k)}= P^{(k)} \cap \NN^{n-1} \, ,
\eq
$$
which are essentially the possible fusion couplings at
level $k$ (by
adding to the elements
of ${\cal S}^{(k)}$ an extra component  $x_0$ equal
to $k$, we recover the usual
fusion couplings).  Because $0< w_i< \infty $, the
transformation
$$
\lambda_i \to \lambda_i'=\lambda_i (w_i/k) \, ,
\quad\qquad
{\bf{v}}_i \to {\bf{v}}_i'=(k/w_i){\bf{v}}_i
\, ,
\quad\qquad w_i \to w_i'=(k/w_i)w_i=k \, ,
\eqlabel\transfo
$$
is well defined, and if we further set
$$
V'= \pmatrix {w_1' & \cdots & w_m' \cr {\bf{v}}_1' &
\cdots & {\bf{v}}_m' \cr}
\, = \, \pmatrix {k & \cdots & k \cr {\bf{v}}_1' &
\cdots & {\bf{v}}_m'\cr}
\, ,
\eqlabel\newV
$$
$P^{(k)}$ can now be given as
$$
P^{(k)} \, = \, \Big \{ {\bf{x}}= \sum_i {\bf{v}}_i'
\, \lambda_i' \,
\Big | \, \lambda' \in
{\RR}^{m}_+ \, , \, \sum_i \lambda_i'=1 \, \Big \}
\, ,
\eqlabel\polyver
$$
which by definition is
the {\it polytope} given by the convex hull of the
vertices ${\bf{v}}_i'$,
$i=1,...,m$.

The main theorem of polytope theory [\ref{ M. Henk,
J. Richter-Gebert, and G. M.  Ziegler, {\it Basic
properties of convex
polytopes},
in {\it Handbook of Discrete and Computational
Geometry}, J.E. Goodman and J.
O'Rourke eds.,  CRC Press 1997.}\refname\Zieg] tells
us that $P$ can
be equivalently described
as a solution set of a finite system of $\ell$ linear
inequalities
(facets), that is
$$
P^{(k)}\, = \, \Big \{ {\bf{x}} \, \Big | \, C \,
{\bf{x}} \leq -k \, b \,
\Big \}
     \,
=   \, \Bigg \{ {\bf{x}} \, \Big | \, C' \,
\pmatrix{k \cr {\bf{x}}} \leq 0 \,
\Bigg \} \, ,
\eqlabel\polyface
$$
where $C'$ is the concatenation $\left(b \, C
\right)$ of $b$ and $C$,
with $C \in  {\RR}^{\ell\times (n-1)}$ and $b  \in
{\RR}^{\ell\times 1}$.
In the last expression, we have made explicit the
fact that the polytope
inequalities translate into  inequalities that the
fusion elements at
$x_0=k$ must satisfy.
We have thus
obtained that
the fusion inequalities are the facets of a polytope.


{}From a practical point of view, we stress that there exist
powerful programs that
give the polytope facets from its vertices (and vice-versa).
The authors
have used the ``cdd'' package of K. Fukuda [\ref{K.
Fukuda, {\it cdd package},
available at
www.ifor.math.ethz.ch/\~{}fukuda.}\refname\Fuku]
for computations in this article.  For a description
of other methods, we refer the reader to [\Zieg].

\newsec{The symmetry group of the fusion polytopes:
generalities }


     The fusion polytope is a geometrical object and so it
is
natural to look for  its symmetry group.  However, for
a polytope there are several different kinds of symmetry
we can consider. For example, consider the polytope $E$
in $\RR^2$ with vertices (0,0), (0,2), (1,2) and (1,0) which
we label 1, 2, 3 and 4.
This polytope is fixed by   the reflections in the lines $y=1$
and $x=1/2$ and by a 180 degree rotation about the point
$(1/2,1)$.
These are examples of  Euclidean (length preserving)
symmetries. Except for the identity transformation,
there are no other Euclidean symmetries of $E$. But there
are additional symmetries if we consider affine transformations,
for example $x \rightarrow 1 -y/2$, $y\rightarrow 2-2x$ fixes
the vertices 2 and 4, while exchanging 1 and 3.

In general, affine  symmetries of a polytope $E$
in $\RR^{n-1}$
are transformations
of the form
$$
{\bar{\zeta}} \, : \, {\bf{x}} \to A \, {\bf{x}} + k
\, b \, ,
\eqlabel\affine
$$
with $A \in  {\RR}^{(n-1)\times (n-1)}$ and $b \in
{\RR}^{(n-1)\times 1}$,
such that
$$
\bar{\zeta} E = E
\eqlabel\polytoinv
$$
If the matrix $A$ is orthogonal then the affine symmetry
is also Euclidean.

A convenient method of finding the affine symmetries
is to identify $\RR^{n-1}$ with the plane $x_0=k$
in $\RR^n$. Then
the affine extension of ${\bar{\zeta}} $, denoted
$\zeta $, is given by
$$
\zeta  \, : \, \pmatrix{ k \cr {\bf{x}} \cr } \to
\pmatrix{  k   \cr   \, A\,{\bf{x}} + k \, b  \cr }
\, ,
\eqlabel\affinevec
$$
or, in matrix form, as
$$
\zeta  \, : \, \pmatrix{ k \cr {\bf{x}} \cr } \to
\pmatrix{ 1 & 0  \cr b &  A   \cr } \,  \pmatrix{ k
\cr {\bf{x}} \cr } \, .
\eqlabel \matrixform
$$

Now, since any polytope is completely characterized by
its vertices or
its facets, a transformation
that leaves $E$ invariant has to leave its set
of vertices and
its set of facets invariant.   That is, we must have
from (\polyver) and
(\polyface), that
the action of ${\bar{\zeta}} $
permutes ${\bf{v}}_1' , \cdots , {\bf{v}}_m'$, so that
$$
{\bar{\zeta}}  \, \pmatrix {{\bf{v}}_1' & \cdots &
{\bf{v}}_m' \cr}=
\bigl( \matrix { \bar \zeta ({\bf{v}}_1') & \cdots &
\bar \zeta (
{\bf{v}}_m' ) \cr} \bigr) =
     \pmatrix {{\bf{v}}_1' & \cdots & {\bf{v}}_m' \cr}
{\sigma}
\, ,
\eqlabel\permuface
$$
for some permutation matrix
$\sigma$.
Using $V'$, whose columns
are vectors of the type $\pmatrix{k \cr {\bf{x}}
\cr}$, the requirement
(\permuface) can be put in matrix form:
$$
B(\zeta)\, V' = V'{\sigma}
\eqlabel\finaleq
$$
where $B(\zeta)$ is the
matrix of the transformation $\zeta$ defined in (\matrixform).
We shall be mainly interested in finding the affine symmetries
of the fusion polytopes and methods for finding these symmetries
will be described in the next section.

Note, however, that in general potential symmetries of a
polytope can often be ruled out
by considering the polytope's combinatorial structure.
Consider, for example, the polytope $E$ introduced at
the beginning of this section.
As mentioned above, a
symmetry of $E$ must permute vertices and facets
and so there cannot be
a symmetry of any type which exchanges vertices 2 and 3 while fixing vertices
1 and 4, since vertices 1 and 2 are joined by a common
edge while 1 and 3 are not.

     One type of combinatorial symmetry would be
to consider all permutations of the faces of the polyhedron
which preserve the face lattice (see for
example [\ref{G.M. Ziegler, {\it Lectures
on Polytopes}, Springer-Verlag,
1995.}\refname\Ziegler] page 128]). However, a simpler type
of combinatorial symmetry, which is easy to find in our examples,
is to look for permutations of vertices and facets which
preserve the vertex-facet incidence matrix. This matrix is easy
to calculate given the vertices and the inequalities representing
the facets: the incidence matrix has an entry 1 in position $(i,j)$
if the $i$th vertex saturates (i.e.~satisfies with equality) the
$j$th facet inequality and is zero otherwise.

If we label  the
edges 1-2, 2-3, 3-4 and 4-1 of our example polytope
$E$ as 1, 2, 3 and 4, corresponding respectively to the inequalities
$x\geq 0\,,y\leq 2\,,x\leq 1\,,y\geq 0$,  we obtain the
incidence matrix:
\def\I{{\cal I}}

$$
\I= \pmatrix{ 1&0&0&1\cr
                1&1&0&0\cr
                0&1&1&0\cr
                0&0&1&1\cr
}\eq$$

If
permutations $\sigma$ and $\tau$ of the vertices and the edges
respectively satisfy $\sigma \I \tau = \I$ then we say that we have
an incidence symmetry. If $\sigma$ and $\tau$ are such
symmetries then $ \sigma \I \tau \tau^{-1} \I^t \sigma^{-1}
= \I \I^t$ and so $\sigma$ commutes with $\I \I^t$. Call
any vertex permutation which commutes with $\I \I^t$ a
vertex symmetry.

Note that any affine symmetry induces
a combinatorial face lattice symmetry and also a combinatorial
incidence symmetry and a vertex symmetry.
These maps are injective since if an
affine symmetry fixes all the vertices of a polytope then
it fixes the whole polytope since a polytope is the convex
hull of its vertices. Thus we have the inclusions:
Euclidean symmetries $\subseteq$ affine symmetries $\subseteq$
combinatorial face lattice symmetries $\subseteq$ combinatorial
incidence symmetries $\subseteq$ vertex symmetries. Thus finding
the vertex symmetries gives a ``upper bound'' on the other types
of symmetries.

For the polytope $E$ we have
$$\I \I^t = \pmatrix{ 2&1&0&1\cr
                       1&2&1&0\cr
                       0&1&2&1\cr
                       1&0&1&2\cr}\eq
$$
Since any vertex symmetry $\sigma$ commutes with $\I \I^t$, it
must also map eigenvectors to eigenvectors and using this it
is not difficult to show that every vertex symmetry of $E$
arises from an affine symmetry.

     In general, however, the group of combinatorial symmetries will be
much larger than the group of affine symmetries. For example
if we take a ``generic'' convex $n$-gon in $\RR^2$, its combinatorial
symmetries
will be a dihedral group, but there will be no affine symmetries
except for the identity transformation.

     Perhaps surprisingly, we find that for our examples that
every vertex symmetry comes from an affine symmetry and so
the most economical method of finding the affine symmetries
would be to calculate first the vertex symmetries. However, this
is not the usual situation and the only real advantage in our
cases turns out
to be that the matrices we have to calculate have smaller entries
if we use vertex symmetries. So in the next section we
give general methods for finding all the affine symmetries
for any polytope.

In the rest of this paper we shall only be concerned with finding
affine symmetries of our polytopes, and so, unless stated otherwise,
by a symmetry  we shall mean an affine symmetry. Similarly when we
refer to a vertex permutation as being a symmetry we mean that
it arises from an affine symmetry as explained at the start of
the next section.


\newsec{The symmetry group of the fusion polytopes:
technical tools }

In this section we will introduce some tools for finding
the affine symmetries of our fusion polytopes introduced
in the last section. We do this by considering vertex permutations
and finding which ones arise from affine symmetries.
These tools, however,
apply in more generality than these applications and so we
start with a more general definition of a symmetry of a matrix:

\proclaim Definition. Let $M$ be an $n\times m$ matrix. An $m\times m$
permutation matrix $\sigma$ is a symmetry of $M$
iff $M\sigma = BM$ for some $n\times n$ matrix
$B$.

The set of symmetries of $M\in\NN^{n\times m}$
is a subset of the group of permutation matrices which
is closed under multiplication, hence
is a group. The corresponding set of
matrices
$B$ does not necessarily form a group, but it does
if the rank of $M$ is
$n$.

This definition is inspired by equation (\finaleq)
since
for the fusions of the affine Lie algebra
$\gh$ we will take $M$ to be the matrix $V'$.
    So
     in the terminology of the last section,
a symmetry of $V'$ is a vertex symmetry
of our fusion polytope which comes from an
affine symmetry. As explained in the
last section we will call these vertex symmetries simply
symmetries to avoid tedious repetition. The group of symmetries of $V'$
will be
denoted by $G[{\gh}]$.  For all our examples
the rank of $V'$ turns out to be $n$, in other
words the fusion polytope has ``full-dimension".

\proclaim Proposition 1.  An $m\times m$
permutation matrix $\sigma$ is a symmetry of the
$n\times m$ matrix $M$
iff  $\sigma N(M) =N(M)$ where $N(M) = \{z\mid
Mz=0\}$.

\noindent{\bf Proof:}
If $z\in N(M)$ then
$M\sigma z =  B Mz = 0$, so $\sigma
N(M)\subseteq N(M)$.
Also $\sigma ^{-1}N(M)\subseteq N(M)$, since
$\sigma^{-1}$ is also a symmetry
of $M$ (because the symmetries form a group) and so
$\sigma N(M) =N(M)$.
Conversely, if $\sigma N(M) = N(M)$
then the matrices $M$ and $M\sigma$ have the same
null space. Thus
their rows generate the annihilator of $N(M)$. In
particular the
rows of $M\sigma$ are contained in the span of the
rows
of $M$ and so there is a
matrix $B$ such that $B M = M\sigma$. So
$\sigma$ is a symmetry of $M$.

When the kernel of $M$ has dimension 0 or 1, this
proposition is sufficient to classify the symmetries.
As we will show in the following sections, this
holds for the fusion polytopes for
$\su(2)$ and $\su(3)$.

If the kernel has larger dimension, we will need another approach.
However, to do so we make the additional assumptions
that $M$ has real entries and that its rank is $n$.
As noted above, the fusion polytope matrices $V'$
have these properties.

With these two assumptions, using the Gram-Schmidt
procedure, we can find an invertible $n\times n$
matrix
$L$ such that the rows of $W=LM$ form an orthonormal
basis of the row space of $M$.  Since $M\sigma = B M$
if and only if $W\sigma = LBL^{-1} W$, it is clear that
$\sigma$ is a symmetry of $M$ if and only if $\sigma$
is a symmetry of $W$.

\proclaim Proposition 2. $\sigma$ is a symmetry of
$M$ if and only if
$Q\sigma = \sigma
Q$, where $Q=  W^\top W.$


\noindent{\bf Proof:} If $\sigma$ is a symmetry of $M$ then
$\sigma$ is a symmetry of $W$ and so
$W\sigma = TW$ for some $n\times n$ matrix $T$.
Moreover, since $\sigma$ acts as an orthogonal
transformation
on the rows of $W$, $T$ is an orthogonal matrix.
Hence,
$W\sigma = TW$ implies $\sigma^{-1}W^\top = W^\top
T^{-1}$.
So
$Q\sigma = W^\top W\sigma = W^\top T W =
\sigma W^\top W$ as required.
For the converse,
note that the matrix $Q$   performs the orthogonal
projection onto the row space of $M$.
So if $\sigma$ commutes with the projection
matrix $Q$ then it maps the row space of $M$ to
itself and so is a symmetry of $M$.

Since $\sigma$
is a permutation, we can read off some of its
properties directly from $Q$. For example a row of
$Q$ can be mapped to another row of $Q$ by a symmetry
only if the two rows have the same set of entries.
A second simplification occurs by observing
that if $\sigma$ is a symmetry and $\sigma u=
\lambda u$ for some vector $u$ then $\sigma Qu =
Q\sigma u
= \lambda Qu$. In particular, if $u$ is fixed
by the symmetry group, so is $Qu$. We will apply these
two observations when we compute the symmetries for
the $\sp(4)$ fusion polytope.


\newsec{The symmetry group of the $\su(2)_k$ polytope}

For $\su(2)$, the matrix $V'$ takes the form
$kV$ with $V$ given in (\vdeu); $N(V')$ is thus
trivial.
Therefore, every permutation
is a symmetry. Since $n=m=4$, this gives $S_4$ as the
symmetry group
$G[{\su(2)}]$.
$S_4$ is generated by the permutations $(1,2,3,4)$
and $(1,2)$, where
$(i,j,\cdots, k)$ stands for a
cyclic permutation of
$i,j,\cdots,k$.  Since
$V'$ is invertible, we can easily find the
corresponding transformation $\zeta$
acting on
$x^\top=\pmatrix{k,\lambda_1,n_{11},n_{12}}$.  It
reads:
$$
\eqalign{
(1,2,3,4): (k,\lambda_1,n_{11},n_{12}) &\mapsto
\pmatrix{ k,  k-\lambda_1-n_{11}+n_{12},  \lambda_1-n_{12},
    k-\lambda_1-n_{11} \cr}  \cr
(1,2):(k,\lambda_1,n_{11},n_{12}) &\mapsto
\pmatrix{
k,  k-n_{11}-n_{12},  n_{11},  k-\lambda_1-n_{11} \cr
}  \cr
}\eq$$

The fusion basis is given in (\inedeux).
Labeling these inequalities from 1 to 4, the
symmetries $(1,\, 2,\, 3,\, 4)$
and $(1,\,2)$ permute the inequalities.
However, the inequalities correspond to
the polytope facets, which being dual to
the vertices transform by the inverse
of the  vertex transform.
Thus if $x\mapsto Bx$ is the transformation
corresponding to
the vertices, then $\alpha^{\top} \mapsto \alpha^{\top}B^{-1}$
is the
appropriate transformation for the facets. The
necessity of this can be seen from the fact that
the transformations should preserve the incidence
relations of the vertices and facets and that this
involves quantities of the form $\alpha^{\top}x$.

{}From this it follows that
the action of $(1,\, 2)$ is $[[1,\, 2],\,[3],\,[4]]$,
i.e., it permutes
the first two inequalities and fixes the last two.
Similarly
the action of $(1,\,2,\,3,\,4)$ is
$[[1,\,2,\,3,\,4]]$.
Thus the fusion basis is generated by any one
of its inequalities under that action of the
symmetry group.

Is there a simple way to understand these symmetries
in terms of the symmetries of the fusion coefficients?
First, notice that the fusion
coefficients are described in terms of a smaller number of
labels than those necessary for the complete
description of the fusion basis.
The complete set of labels can be split into  two subsets:
the  Dynkin labels of the three weights under
consideration and the
`missing labels'.
If some  symmetries do not
involve in an essential way the
missing labels, they will project onto
fusion-coefficient symmetries. However, if the
missing labels are an essential part of the symmetry
transformations, the symmetry will disappear in the
projection. For instance, tensor-product coefficients
can be obtained by projection of the fusion
coefficients. The latter require an extra variable for their
description, the level $k$, and fusion
coefficients have more symmetries than the tensor-product
coefficients. The extra symmetries are
the outer automorphisms -- see the next paragraph -- and these
involve the level in an essential way.
Let us make the general situation more precise: Denote collectively the finite Dynkin
labels $\{\la_i,\,\mu_i,\,\nu_i\}$ and
$k$ by $D$ and the set of missing labels by $\gamma$. A facet
symmetry is generically of the form
$\{D,\gamma\}\rw \{D'(D, \gamma),\gamma'(D, \gamma)\}$. This will be
a symmetry of the fusion coefficients
only when $D'$ does not depend upon $\gamma$.
Therefore there is no simple relationship
between
     the
symmetries of the  facets and the
symmetries of the fusion
coefficients.   In
this regard, each algebra has to be studied separately.

The symmetries of the  fusion coefficients
include those that are level-independent, i.e., the
symmetries of the
corresponding tensor-product coefficients; these are
the conjugation of the
three
weights, $(\lah,\muh,\nuh) \rw
(\lah^*,\muh^*,\,\nuh^*)$ and the different
permutations of $\lah,\muh$ and $\nuh^*$.  The
remaining
symmetries are intrinsically affine.  These  include
the outer-automorphism
symmetries which
   take the following form: if
$A,A'$ are two arbitrary elements of the outer-
automorphism group of $\gh$, the
fusion coefficients satisfy
$$ {\Nc_{A\lah, A'\muh}^{(k)}}^{~AA'\nuh}=
{\Nc_{\lah\muh}^{(k)}}^{~\nuh}\eq$$ For $\su(N)$,
this group has
order $N$: $A^N=1$. The symmetry group can
be larger than that generated by tensor product symmetries
and outer automorphisms, see for example [\ref{D. Verstegen,
{\it New Exceptional Modular Invariant Partition Functions For Simple 
Kac-Moody Algebras},
Nucl.\ Phys.\ B {\bf 346}, 349 (1990).}\refname\V,\ref{
J. Fuchs, A. N. Schellekens and C. Schweigert,
{\it Galois modular invariants of WZW models},
Nucl.\ Phys.\ B {\bf 437}, 667 (1995).}\refname\FSS,\ref{
T. Gannon,
{\it The automorphisms of affine fusion rings},
arXiv:math.qa/0002044.
}\refname\G].
There are often symmetries which exist for some, but not all levels.
The method we present here will not detect this type of symmetry and from this
point we will exclude them. In other words, by symmetries of
the fusion coefficients we mean symmetries which exist for all levels.

For $\su(2)$, there are no missing labels. Hence the
symmetry group of the
polytope
must be identical to the symmetry group of the fusion
coefficients which exist for all $k$ and that
leave $k$
fixed.
This group is isomorphic to the semi-direct product
$(S_2\times S_2):S_3$.
The $S_3$ factor
comes from  the permutation of the three weights,
while the two factors of $S_2$
account for the two copies of the outer automorphism
group (one acting on the
weight $\lah$, the other acting on
$\muh$).  The conjugation action of $S_3$ is via
the outer automorphisms of $S_2\times S_2$ which
permute the non-identity elements. The group
$S_4$ contains a group $S_2\times S_2$ generated
by the cycles of type $2^2$. Any of the four $S_3$ subgroups
act on this $S_2\times S_2$ as outer automorphisms by
conjugation and so
     the symmetry  group is isomorphic to
$S_4$.


Let us first reexpress all the basis symmetries in
terms of the  Dynkin labels:
$$\eqalignD{
(1,2,3,4) : & (k, \lambda_1, \mu_1,
    \nu_1) \rightarrow (k,k-\nu_1, k-\mu_1, \lambda_1) \cr
(1,2) : &  (k, \lambda_1, \mu_1, \nu_1) \rightarrow
(k, k-\mu_1,
k-\lambda_1, \nu_1) \cr
(2,3) : & (k, \lambda_1, \mu_1, \nu_1) \rightarrow
(k,\lambda_1,\nu_1,
\mu_1) \cr
(3,4) : & (k, \lambda_1, \mu_1, \nu_1) \rightarrow
(k,\mu_1,\lambda_1, \nu_1)
\cr} \eq$$
(Clearly, the last two symmetries can be obtained
from the first two.)

Let us make explicit the correspondence between these
symmetries and
symmetries of
the fusion coefficients. For this, notice first that
the multiplicity of the
$\su(2)_k$ product $\hat{\lambda} \times
\hat{\mu} \supset \hat{\nu}$ is the same as that of
$\hat{\lambda} \times
\hat{\mu}\times \hat{\nu} \supset 0$. Let $P_{12}$ be
the operator that permutes
the first two weights and similarly for the other
permutation operators.
Moreover, let $a$ be the
$\su(2)$ automorphism that interchanges the two
simple roots, hence the two
Dynkin
labels: $a[k-\la_1,\la_1]= [\la_1,k-\la_1]$.
Therefore, the nontrivial
actions on
$\lah\times \muh\times \hat{\nu} \supset 0$ are
simply:
$$
     a \lah\times a \muh\times \hat{\nu} \supset
0\qquad\qquad
     a \lah\times \muh\times a\hat{\nu} \supset
0\qquad\qquad
     \lah\times a \muh\times a\hat{\nu} \supset 0\eq$$
Denote these actions respectively as
$(a,a,1),\;(a,1,a),\;(1,a,a)$.
The fusion basis symmetries can then be related
directly to $a$ and $P$ actions
as follows:
$$\eqalignD{
(1,2,3,4) &\equiv (a,a,1) P_{13}\cr
(1,2) &\equiv (a,a,1) P_{12}\cr
(2,3) & \equiv P_{23}\cr
(3,4) & \equiv P_{12}\cr}\eq
$$
`Pure' finite or affine symmetries can be obtained by
composition, e.g.,
$$\eqalignD{
    (2,3,4)  &  \equiv P_{321} \cr
(1,2)(3,4)  & \equiv (a,a,1) \cr}\eq $$

\newsec{The symmetry group of the $\su(3)_k$ polytope
}

The situation for $\su(2)$ is not typical in two
ways. First, there
are no missing
labels; hence any permutation of the vertices is
bound to be a
symmetry of the fusion coefficients.  In addition,
there are no linear relations
between the elementary solutions.  Such relations
will induce severe constraints
on the possible lifts of the fusion-coefficient
symmetries to
polytope symmetries.

For $\su(3)$, the elementary couplings are (using the
notation
$\lah = [\la_0,\la_1,\la_2]$ with
$\la_0+\la_1+\la_2=k$ and the LR
variables - cf.
[\BCM]. Note that $\la_1$ and $\la_2$
are Dynkin labels and not the partition labels
to which they usually refer in the Littlewood-Richardson rule
   as in [\Mac] chapter 1 for example):
$$\eqalignT{
\E_0: \quad[1,0,0]\times [1,0,0]\supset [1,0,0]:&
\quad d \quad
&(1,0,0,0,0,0,0,0)\quad & (1,1)
\cr
\E_1:\quad[0,1,0]\times [0,0,1]\supset [1,0,0]: &
\quad
dL_1N_{12}N_{23}\quad &(1,1,0,0,1,0,0,1) \quad &
(a,a^2)\cr
\E_2:\quad[0,1,0]\times [1,0,0]\supset [0,1,0]: &
\quad
dL_1\quad &(1,1,0,0,0,0,0,0) \quad & (a,1)\cr
\E_3:\quad[1,0,0]\times [0,1,0]\supset [0,1,0]: &
\quad
dN_{11}\quad &(1,0,0,1,0,0,0,0) \quad & (1,a)\cr
\E_4:\quad[0,0,1]\times [0,1,0]\supset [1,0,0]: &
\quad
     dL_2N_{13}\quad  &(1,0,1,0,0,1,0,0) \quad &
(a^2,a)\cr
\E_5:\quad[0,0,1]\times [1,0,0]\supset [0,0,1]: &
\quad
d L_2\quad &(1,0,1,0,0,0,0,0)\quad & (a^2,1) \cr
\E_6:\quad[1,0,0]\times [0,0,1]\supset [0,0,1]: &
\quad
dN_{11}N_{22}\quad &(1,0,0,1,0,0,1,0)\quad & (1,a^2)
\cr
\E_7:\quad[0,1,0]\times [0,1,0]\supset [0,0,1]: &
\quad
d L_1N_{12}\quad &(1,1,0,0,1,0,0,0) \quad & (a,a)\cr
\E_8:\quad[0,0,1]\times [0,0,1]\supset [0,1,0]: &
\quad
d L_2N_{11}N_{23}\quad &(1,0,1,1,0,0,0,1)\quad &
(a^2,a^2) \cr }\eq$$
Besides each coupling, we have written  the
`exponential description',
the corresponding  vector $\e_i$ with entries in the
order
$ (k,\la_1,
\la_2, n_{11}, n_{12}, n_{13}, n_{22}, n_{23})$, as
well as the action
of the outer automorphism on the first two weights of
$\E_0$ (in the form
$(a^n,a^m)$) that yields the coupling under
consideration.

The corresponding facets (the fusion basis) are found
to be
$$\eqalignD{
n_{12}&\geq 0\qquad\qquad &\la_2+n_{12}-n_{13}-
n_{23}\geq 0\cr
n_{13}&\geq 0\qquad\qquad &n_{11}-n_{22}\geq 0\cr
n_{22}&\geq 0\qquad\qquad &n_{11}+n_{12}-n_{22}-
n_{23}\geq 0\cr
n_{23}&\geq 0\qquad\qquad &k-\la_1-\la_2-n_{22}\geq
0\cr
\la_1-n_{12}&\geq 0\qquad\qquad &k-\la_1-\la_2-
n_{11}+n_{23}\geq 0\cr
\la_2-n_{13}&\geq 0\qquad\qquad & k-\la_1-n_{13}-
n_{11}\geq 0\cr}
\eqlabel\inee$$
This agrees with the system of inequalities obtained
in [\BCM].

For $\su(3)$, the matrix $V'$ is $8\times 9$:
$$V' = k\pmatrix{
1 & 1 & 1 & 1 & 1 & 1 & 1 & 1 & 1 \cr
0 & 1 & 1 & 0 & 0 & 0 & 0 & 1 & 0 \cr
0 & 0 & 0 & 0 & 1 & 1 & 0 & 0 & 1 \cr
0 & 0 & 0 & 1 & 0 & 0 & 1 & 0 & 1 \cr
0 & 1 & 0 & 0 & 0 & 0 & 0 & 1 & 0 \cr
0 & 0 & 0 & 0 & 1 & 0 & 0 & 0 & 0 \cr
0 & 0 & 0 & 0 & 0 & 0 & 1 & 0 & 0 \cr
0 & 1 & 0 & 0 & 0 & 0 & 0 & 0 & 1 \cr
}\eq$$
and since the rank of $V'$ is 8, $N(V')$ is one-
dimensional
and is spanned by
$$
w^\top=
\pmatrix{
1 & -1 & 0 & -1 & 0 & -1 & 0 & 1 & 1 \cr
}\eq$$

The condition on $\sigma$
is thus $\sigma w = \lambda w$ for some scalar
$\lambda$ and,
since the eigenvalues of $\sigma$ are roots of unity,
$\lambda$
is either 1 or $-1$.
Thus $G[{\su(3)}]$ is isomorphic to $S_3\times
(S_3\wr  S_2)$.
The  $S_3$  permutes  the 0's of $w$
and $S_3 \wr S_2$ (the wreath product of $S_3$ and
$S_2$
i.e., the direct product of {\it two} $S_3$'s with
$S_2$ acting by exchanging
them)
permutes and exchanges the 1's and  $-1$'s.

Let $x^{\top} = (k,\la_1,
\la_2, n_{11}, n_{12}, n_{13}, n_{22}, n_{23})$.
The action of a set of generators of
the symmetry group  on the variables is:
$$
\eqalign{
(3,5,7):x^{\top}& \mapsto
     (k, n_{12}+n_{22}, \lambda_1+\lambda_2-n_{12}-n_{13},
n_{11}+n_{13}-n_{22}, n_{12},\lambda_1- n_{12},
n_{13}, n_{23})\cr
(3,5):x^{\top}& \mapsto
(k, n_{12}+n_{13}, \lambda_1+\lambda_2-n_{12}-n_{13},
n_{11}, n_{12}, \lambda_1-n_{12}, n_{22}, n_{23})\cr
(1,8,9):x^{\top}& \mapsto
(k, k-\lambda_2-n_{11}-n_{12}+n_{23}, \lambda_2+n_{12}-n_{23},
n_{11}+n_{12}-n_{23},\cr &\qquad\; k-\lambda_1-\lambda_2-n_{11}+n_{23},
n_{13}, n_{22}, n_{12}) \cr
(1,8):x^{\top}&
\mapsto
(k, k-\lambda_2-n_{11}-n_{12}+n_{23}, \lambda_2,
n_{11},
     k-\lambda_1-\lambda_2-n_{11}+n_{23}, n_{13}, n_{22},
n_{23})
\cr}\eq$$ together with
$$
\eqalign{(1,2)(4,8)(6,9):x^{\top}
\mapsto
(&k, k-\lambda_2-n_{12}- n_{22}, \lambda_2,
\lambda_2+n_{12}-n_{13}+n_{22}-n_{23},\cr & k-
\lambda_1-\lambda_2-n_{22},
n_{13},
n_{22}, k-\lambda_1-n_{13}-n_{11}) \cr }\eq
$$

The fusion basis is given by (\inee). Labeling these
inequalities successively,
column by column, from 1 to 12 (i.e., 8 corresponds to
$n_{11}-n_{22}\geq 0$), we find
that the action of the symmetry group on the
inequalities considered as polytope
facets is generated by:
$$\eqalign{
(3,5,7)&:[[1], [2, 3, 5], [4], [6], [7], [8], [9],
[10], [11], [12]] \cr
(3,5)&: [[1], [2, 5], [3], [4], [6], [7], [8], [9],
[10], [11], [12]]  \cr
(1,8,9)&: [[1, 4, 11], [2], [3], [5], [6, 12, 7], [8,
10, 9]] \cr
(1,8)&:  [[1, 11], [2], [3], [4], [5], [6], [7, 12],
[8], [9, 10]]  \cr
(1,2)(4,8)(6,9)&: [[1, 10], [2], [3], [4, 12], [5],
[6], [7, 8], [9], [11]]  \cr
}\eq$$
     where as before
$[i,j,\cdots, k]$ stands for a
cyclic permutation of the inequalities
$i,j,\cdots,k$.
Thus there are two orbits under the symmetry group.
One consisting of the inequalities $2,\, 3$ and 5
and the other consisting of the inequalities
$1,\, 4,\, 6,\, 7,\, 8,\, 9,\, 10,\, 11$ and 12. So
the fusion basis
is generated by the two inequalities
$n_{12}\geq0$ and $n_{13}\geq 0$ under the action
of $G[{\su(3)}]$.

Let us now try to understand these results.  As
already pointed out,
these symmetries must
be compatible with the symmetries of the
fusion
coefficients.  But there is a further constraints on
the fusion symmetries that has not
been spelled out yet: in general there are linear
relations among the elementary
couplings and the symmetries of the facets must
preserve these relations.
In the
$\su(3)$ case, there is only one such linear
relation, which is:
$$\E_0\E_7\E_8=\E_1\E_3\E_5\eq$$
This explains the existence of the three $S_3$ blocks:
permutations
among the sets $\{\E_0,\E_7,\E_8\}$ and
$\{\E_1,\E_3,\E_5\}$ are symmetries that
preserve each sides of the relation.  The third $S_3$
factor is bound to relate
the three remaining vertices.  Moreover, the $S_2$
group generated by
$(1,2)(4,8)(6,9)$ is another transformation that
leaves the
relation unchanged: but instead of leaving each side
invariant, it interchanges
the two sides of the relation: $\E_0\lra \E_1,
\;\E_3\lra \E_7, \;\E_5\lra
\E_8$. The occurrence of a linear relation is thus
responsible for the fact that the
symmetry group is not $S_9$ but only a subgroup
thereof.

%
Of course, the existence of relations is intimately connected with
the matrix $V'$. In fact,
$N(V')$ is precisely  the set of generating
relations (in the sense of [\BCM]) out of which all the relations can
be obtained.

To complete the analysis of the $\su(3)_k$ polytope
symmetries, let us
investigate their explicit relation with symmetries of the
fusion coefficients. For
this, we first reexpress the symmetry transformations
in terms of the Dynkin
labels of the three weights plus
$n_{23}$.  Thus $n_{23}$ is  the missing label. We also reformulate
the results in terms of
the symmetrized product:
     $\hat{\lambda} \times
\hat{\mu}\times \hat{\nu} \supset 0$. Recall that
the multiplicity of  $\hat{\lambda} \times
\hat{\mu} \supset \hat{\nu}^*$ is the same as that of
$\hat{\lambda} \times
\hat{\mu}\times \hat{\nu} \supset 0$, where $*$
denotes the finite weight
conjugation; for $\su(3)$, it amounts to interchanging
the two finite Dynkin
labels.
The precise transformation relations are as follows:
$$\eqalignD{
& n_{11}=L_2-\lambda_1-\lambda_2\qquad
& n_{12}=L_1-L_2+n_{23}\cr
& n_{13}=L_2-\nu_1-\nu_2-n_{23}\qquad
& n_{22}=\mu_2-n_{23}\cr}\eq$$
    With the
vector
$y^{\top}$ defined as
$y^{\top}=(k,\lambda_1,\lambda_2,\mu_1,\mu_2,
\nu_1,\nu_2, n_{23})
$, we can rewrite the symmetries of the fusion basis
as:
%
%
$$\eqalignD{(3,5,7) : & y^{\top} \rightarrow
(k,L_2-\lambda_2-\nu_2,L_2-\mu_1-\mu_2,L_2-\lambda_2-
\mu_2,L_2-\nu_1-\nu_2
,L_2-\mu_2-\nu_2, \cr
& \quad\quad\quad  L_2-\lambda_1-\lambda_2,
n_{23})\cr
     (3,5) : & y^{\top}
\rightarrow (k, L_1-\nu_1-\nu_2, L_2-\mu_1-\mu_2,
L_2-\lambda_2-\mu_2,\mu_2, \nu_1,
L_1-\lambda_1-\nu_1, n_{23})\cr
(1,8,9) : & y^{\top} \rightarrow
(k,k-L_1+\lambda_1, L_2-\mu_2-\nu_2,
\ k-L_1+\mu_1, L_2-\lambda_2-\nu_2,
\cr
&\quad\quad\quad k-L_1+\nu_1, L_2-\lambda_2-\mu_2, L_1-L_2+n_{23})\cr
    (1,8) : & y^{\top}
\rightarrow (k, k- L_1+\lambda_1, \lambda_2,
k-L_1+\mu_1,\mu_2, k-L_1+\nu_1, \nu_2, n_{23})\cr
(1,2) (4,8) (6,9) : & y^{\top} \rightarrow (k,
k-L_2+\nu_2,\lambda_2,\mu_1,k-L_1+\nu_1,
L_2-\lambda_1-\lambda_2,L_1-\mu_1-\mu_2, \cr
& \quad\quad\quad n_{23}+k-L_1-\mu_2+\nu_1)\cr  }\eq
$$
where $L_i=(\la+\mu+\nu,\omega_i)$ with $\omega_i$
the $i$-th fundamental
weight. The remarkable feature of these symmetry
transformations is that they send
$y^{\top} \rightarrow {y'}^\top$ such that none of
the prime variables except
$n'_{23}$ depends upon $n_{23}$.  In other words, the new
Dynkin labels, collectively denoted by $D'$, do
not depend upon  $n_{23}$. Therefore, these symmetries map a state (i.e.,
a tableau) of a given fusion to another state of
another fusion. The same is
necessarily  true for the inverted transformations.
There is thus a one-to-one
correspondence between the two fusions, i.e., they
have the same multiplicity! In
other words, the symmetries of the fusion basis are
symmetries of the fusion
     coefficients. For instance,
$$ \eqalignD{(3,5,7) : & \quad {\cal N}_{(\lambda_1,
\lambda_2) (\mu_1,\mu_2) (\nu_1,\nu_2)
}^{~ (k) } ={\cal N}_{(L_2-\lambda_2-\nu_2,L_2-\mu_1-
\mu_2) (L_2-\lambda_2-\mu_2,L_2-\nu_1-\nu_2) (L_2-
\mu_2-\nu_2, L_2-\lambda_1-\lambda_2)}^{~ (k)} \cr
(3,5) : & \quad {\cal N}_{(\lambda_1, \lambda_2)
(\mu_1,\mu_2) (\nu_1,\nu_2)
}^{~ (k) } ={\cal N}_{(L_1-\nu_1-\nu_2, L_2-\mu_1-
\mu_2) (L_2-\lambda_2-\mu_2,\mu_2) ( \nu_1,L_1-
\lambda_1-\nu_1)}^{~ (k)} \cr
(1,8,9) : & \quad {\cal N}_{(\lambda_1, \lambda_2)
(\mu_1,\mu_2) (\nu_1,\nu_2)
}^{~ (k) } =
{\cal N}_{(k-L_1+\lambda_1, L_2-\mu_2-\nu_2)
(k-L_1+\mu_1, L_2-\lambda_2-\nu_2)
(k-L_1+\nu_1, L_2-\lambda_2-\nu_2)}^{~ (k)}
\cr (1,8) : & \quad {\cal N}_{(\lambda_1, \lambda_2)
(\mu_1,\mu_2) (\nu_1,\nu_2)
}^{~ (k) } ={\cal N}_{(k-L_1+\lambda_1, \lambda_2) (k-
L_1+\mu_1, \mu_2)
(k-L_1+\nu_1,\nu_2)}^{~ (k)} \cr
(1,2) (4,8) (6,9) : & \quad {\cal N}_{(\lambda_1,
\lambda_2) (\mu_1,\mu_2) (\nu_1,\nu_2)
}^{~ (k) } ={\cal N}_{(k-L_2+\nu_2,\lambda_2)
(\mu_1,k-L_1+\nu_1)
(L_2-\lambda_1-\lambda_2,L_1-\mu_1-\mu_2)}^{~ (k)}
\cr}\eq$$
In this rewriting, the dependence upon the $n_{23}$
variable is dropped.

These are clearly new fusion
symmetries as they mix the labels of the three representations.
It is simple to
verify that they leave the explicit expression of the
$\su(3)_k$ fusion
coefficients (given in [\ref{L. B\'egin, P. Mathieu and M.A.
Walton, Mod. Phys.
Lett. A, Vol. 7
     (1992) 3255.}\refname\LB]) invariant.  Here is a
numerical illustration: $(1,8)$
maps
$$[4,10,2] \times [1,4,11]\times [9,3,4] \supset 0
\leftrightarrow  E_0 E_1^8 E_3^2 E_6 E_7^2 E_8^2,
\quad E_1^8 E_2 E_3^2 E_4 E_6^2 E_7 E_8   \eq$$
onto
$$[5,9,2] \times [2,3,11]\times [10,2,4] \supset 0
\leftrightarrow E_7 E_1^8 E_3^2 E_6 E_0^2 E_8^2,
\quad E_1^8 E_2 E_3^2 E_4 E_6^2 E_0 E_8    \eq$$


\noindent Both fusions have two decompositions in
terms of elementary couplings, meaning that they have the
same multiplicity. This symmetry is manifestly
distinct from those already known. It is not a
trivial symmetry in that, in terms of the
elementary-coupling decomposition
s, it
corresponds to the interchange of $E_0$ and
$E_7$ in each decomposition.

The usual fusion symmetries can be obtained from
various combinations of the
above symmetries of the fusion basis. For instance,
%
%

$$\eqalignD{(2,4,6)(3,5,7) &  \equiv
P_{123} \cr
(8,9) (5,7) (2,4) & \equiv C ~P_{13} \cr
(2,8)(1,4)(6,9) & \equiv (1,a,a^2)
~P_{23}   \cr
\cr}\eq $$
where  $C$ is the  conjugation:$$\ C\{\hat{\lambda}
\times
\hat{\mu}\times \hat{\nu} \supset 0\}=
\hat{\lambda}^* \times
\hat{\mu}^*\times \hat{\nu}^* \supset 0\eq$$
We thus recover the known symmetries and additional
ones.

\newsec{The symmetry group of the $\sp(4)_k$ polytope
}

The $\sp(4)$ fusion rules are most conveniently
described by means of the
variables introduced in [\ref{ A.D. Berenstein and
A.V.  Zelevinsky, J.
Geom. Phys.
{\bf 5} (1989) 453.}\refname\BZin] and used in [\BCM,
\BCMa]
(with a slight change of notation),
namely
$\{k, \la_1,\la_2,\mu_1,\mu_2, r_1,r_2,p,q\}$,
with $p, q$ and $r_i/2$ nonnegative integers. The
Dynkin labels of the third
weight are
$$\nu_1  = r_2-r_1-2p+\la_1+\mu_1\; \qquad
\nu_2  = p-q-r_2+\la_2+\mu_2\eq$$

In terms of the exponentiated variables,
the elementary couplings are $\E_0=d$, together with
$$\eqalignT{
&\A_1= dM_1\quad\qquad \qquad
&\A_2=dL_1\quad \qquad\qquad
&\A_3= dL_1M_1PQ\cr
&\B_1=dM_2\quad
&\B_2= dL_2\quad
&\B_3= dL_2M_2R_1^2R_2^2\cr
&\C_1= dL_2M_1Q \quad
&\C_2= dL_1M_2R_2^2P \quad
&\C_3= dL_1M_1P\cr
&\D_1=d^2L_1^2M_2R_2^2P^2\quad\qquad \qquad
&\D_2= d^2L_2M_1^2R_1^2 \quad\qquad \qquad
&\D_3= d^2L_2M_2R_2^2 \cr}\eq$$
Taking these to be the vertices of a polytope, the
corresponding facets are
found to
be:
$$
\eqalignD{
k-\lambda_1-\lambda_2-\mu_2-r_1/2+r_2 &\geq 0 \qquad
\qquad
&\mu_1-q \geq 0 \cr
k-\lambda_1-\lambda_2-\mu_2+r_2/2 &\geq 0 \qquad
\qquad
&\mu_1-q-r_1+r_2 \geq 0 \cr
k-\lambda_1-\mu_1-\mu_2+p &\geq 0 \qquad \qquad
&\mu_1-p-r_1+r_2 \geq 0 \cr
k-\lambda_1-\lambda_2-\mu_1-\mu_2+p+q+r_1/2 &\geq 0
\qquad \qquad
&\mu_2-r_2/2 \geq 0 \cr
\lambda_1-p &\geq 0 \qquad \qquad
&r_1 \geq 0 \cr
\lambda_2-r_1/2 &\geq 0 \qquad \qquad &r_2 \geq 0 \cr
\lambda_2-r_1/2-q+p &\geq 0 \qquad \qquad &p\geq 0 \cr
\lambda_2-r_2/2-q+p &\geq 0 \qquad \qquad & q \geq 0
\cr
}\eqlabel\inesp
$$
in agreement with the results obtained in [\BCM].

We next analyze the symmetries of the $\sp(4)$ fusion
polytope. The
matrix $V'$, with the column ordering $(\E_0, \A_1,
\cdots, \D_3)$, reads
$$
V'=
k \pmatrix{
1 & 1 & 1 & 1 & 1 & 1 & 1 & 1 & 1 & 1 & 1 & 1 & 1 \cr
0 & 0 & 1 & 1 & 0 & 0 & 0 & 0 & 1 & 1 & 1 & 0 & 0 \cr
0 & 0 & 0 & 0 & 0 & 1 & 1 & 1 & 0 & 0 & 0 & 1/2 &
{1/2} \cr
0 & 1 & 0 & 1 & 0 & 0 & 0 & 1 & 0 & 1 & 0 & 1 & 0 \cr
0 & 0 & 0 & 0 & 1 & 0 & 1 & 0 & 1 & 0 & {1/2} & 0 & {
1/2} \cr
0 & 0 & 0 & 0 & 0 & 0 & 2 & 0 & 0 & 0 & 0 & 1 & 0 \cr
0 & 0 & 0 & 0 & 0 & 0 & 2 & 0 & 2 & 0 & 1 & 0 & 1 \cr
0 & 0 & 0 & 1 & 0 & 0 & 0 & 0 & 1 & 1 & 1 & 0 & 0 \cr
0 & 0 & 0 & 1 & 0 & 0 & 0 & 1 & 0 & 0 & 0 & 0 & 0 \cr
}\eq
$$
Unfortunately for this case the kernel of $V'$
is not one-dimensional.
In fact, $N(V')$ is four-dimensional, which we
understand from the fact that  there are
four generating relations
-- cf.
[\BCM] eq. (5.18). As a result, we must study the
symmetries of $V'$ via the
commutant of the matrix
$Q$ that performs the orthogonal projection onto the
row space of $V'$ (cf.
proposition 2 of sect. 6). This matrix reads
$$
Q={1 \over {108}}
\pmatrix{82&16&0&18&0&10&-2&-18&-14&-20&16&4&16\cr
\noalign{\medskip}
16&67&0&-9&0&-20&-14&9&10&13&-14&28&22\cr
\noalign{\medskip}0
&0&108&0&0&0&0&0&0&0&0&0&0\cr \noalign{\medskip}
18&-9&0&81&0&-18&0&27&0&9&18&0&-18\cr
\noalign{\medskip}0&0&0&0&108&0&0&0&0&0&0&0&0\cr
\noalign{\medskip}10&-20&0
&-18&0&82&-2&18&-14&16&16&4&16\cr
\noalign{\medskip}-2&-14&0&0&0&-2&94&0&10&-14&4&28&4
\cr \noalign{\medskip}-18&9&0&27&0&18&0&81&0&-9&-
18&0&18
\cr
\noalign{\medskip}-14&10&0&0&0&-14&10&0&70&10&28&-
20&28\cr \noalign{\medskip}
-20&13&0&9&0&16&-14&-9&10&67&22&28&-14\cr
\noalign{\medskip}16&-14&0&18&0&16&4&
-18&28&22&40&-8&4\cr
\noalign{\medskip}4&28&0&0&0&4&28&0&-20&28&-8&52&-8\cr
\noalign{\medskip}16&22&0&-18&0&16&4&18&28&-14&4&-
8&40 }
\eqlabel\qsp
$$

As already pointed out in sect. 6, a row of $Q$ can
be mapped to another
row by a symmetry transformation $\sigma$ only if the
two rows have the
same entries. Hence, from the explicit form of
(\qsp), we see that any
symmetry fixes the sets $\{1,6\}$, $\{2,10\}$,
$\{3,5\}$, $\{4,8\}$ $\{11,13\}$, $\{7\}$, $\{9 \}$
and $\{12\}$. It is not difficult to see
that the permutation $(3,5)$ is a symmetry.

We can again examine $Q$ to see that
a symmetry exchanges rows 1 and 6 iff it
exchanges columns 2 and 10. It exchanges
rows 2 and 10 iff it exchanges columns 11 and 13
and  exchanges rows 11 and 13 iff it exchanges
columns 4 and 8.

It is easy to verify that (1,6)(2,10)(11,13)(4,8)
is a symmetry. Thus, the symmetries of $V'$
are
$$G[\sp(4)]= \{(~ ), \,(3,5),
\,(1,6)(2,10)(11,13)(4,8), \,
     (3,5)(1,6)(2,10)(11,13)(4,8)\}\eq$$
isomorphic to $S_2\times S_2$.

With $x^{\top}=(k, \lambda_1, \lambda_2, \mu_1,
\mu_2, r_1, r_2, p, q)$,
the changes of variables corresponding
to the generators of the symmetry group are:
$$\eqalign{
(3, 5) \,:~
x^{\top} \mapsto
(&k, \mu_2-r_2/2+p, \lambda_2, \mu_1, \lambda_1+r_2/2-
p, r_1, r_2, p, q)   \cr
(1, 6)(2, 10)(11, 13)(4, 8)\,:~
x^{\top} \mapsto
(&k, \lambda_1+\mu_1-r_1+r_2-2p,
     k-\lambda_1-\lambda_2-\mu_1-\mu_2+r_1+p+q,\cr
&\mu_1, \mu_2, r_1, r_2,
\mu_1-p-r_1+r_2, q) \cr  }\eq$$

The orbits of the symmetries on the inequalities
of the fusion basis are (with the inequalities
(\inesp) labelled
consecutively,
column by column,  from 1 to 16):
$$\eqalign{
(3, 5):~&
[[5, 12]] \cr
(1, 6)(2, 10)(11, 13)(4, 8):~&
[[1, 7], [2, 8], [4, 6], [11, 15] ] \cr
}\eq$$
where orbits of length 1 have been omitted.

As indicated in the context of the $\su(3)$ analysis,
the most
severe constraints on the symmetries come from the
linear relations. There
is indeed
a large number of relations for $\sp(4)$ [\BCM]:
$$\eqalignT{ &\E_0\C_1 \C_2 = \A_3 \D_3\quad
&\E_0\C_2\C_3= \A_1\D_1\quad &
\E_0\C_3\C_1=
\A_1\A_3\B_2
\cr &\D_1 \D_2=\E_0\B_3\C_3^2\quad & \D_2\D_3=
\E_0\A_1^2\B_2\B_3\quad &
\D_1\D_3=
\E_0\B_2\C_2^2
\cr &\C_1\D_1= \A_3\B_2\C_2\quad &\C_2\D_2=
\A_1\B_3\C_3\quad & \C_3\D_3=
\A_1\B_2\C_2\cr}
\eqlabel\zysp$$
It is not difficult to check that the symmetries
leave these relations
invariant. In
fact, $(3,5)= (\A_2,\B_1)$ (which means the
interchange of $\A_2$ and $\B_1$) and
these two couplings do not appear in the relations.
The other symmetry reads
$(\E_0,
\B_2)\, (\A_1, \C_3)\,(\D_1, \D_3)\,(\A_3, \C_1)$.

With the vector $y^{\top}$ defined as
$y^{\top}=(k,\lambda_1,\lambda_2,\mu_1,\mu_2,
\nu_1,\nu_2,p,q)
$, we can rewrite the symmetries of the fusion basis
as:
$$\eqalignD{& (3,5) :  y^{\top} \rightarrow
(k, -1/2 \lambda_2+1/2 \mu_2+1/2 \nu_2+1/2 p+1/2 q,
\lambda_2, \mu_1, \cr
& \quad\quad\quad \lambda_1+1/2 \lambda_2+1/2 \mu_2-
1/2 \nu_2-1/2 p-1/2 q,-\lambda_1-1/2 \lambda_2+1/2
\mu_2+\nu_1+1/2 \nu_2+1/2 p+1/2 q, \cr
& \quad\quad\quad  \lambda_1+1/2 \lambda_2-1/2
\mu_2+1/2 \nu_2-1/2 p-1/2 q, p,q) \cr
& (1,6) (2,10) (11,13) (4,8) : y^{\top}
\rightarrow (k,\nu_1, k-\nu_1-\nu_2, \mu_1,\mu_2,
\lambda_1,k-\lambda_1-\lambda_2,-
\lambda_1+\nu_1+p,q)\cr
}\eq $$
The first polytope symmetry does not correspond to a
fusion-coefficient symmetry.
However,  the second one is a  combination of an outer
automorphism and a permutation of two weights:
$(1,6) (2,10) (11,13) (4,8)=(a,1,a) P_{13}$. This is
not a new symmetry of fusion coefficients.
\newsec{The symmetry group of the $\su(4)_k$ polytope
}

The whole set of $\su(4)$ fusion elementary couplings
can all be generated
from  two couplings that have no elementary finite
relative:
$$ \E_0=[1,0,0,0]\times[1,0,0,0]\supset[1,0,0,0]\qquad
\F=[0,1,0,1]\times[0,1,0,1]\supset[0,1,0,1]\eq$$
by means of the outer-automorphism group.  We can thus characterize a
coupling by a pair $(a^n,a^m)_i$ where $a^n$ and $a^m$ act on the first and
the second weight respectively,  understanding that the action on the third
weight is $a^{n+m}$. Here $a$ permutes the Dynkin labels of an affine $\su(4)$
weight as $a[\la_0,\la_1,\la_2,\la_3]=[\la_3,\la_0,\la_1,\la_2]$
     so that $a^4=1$.
The subindex $i$ refers to the elementary coupling $\E_0$ or $\F$ from which it
is obtained; these are labelled respectively as $i=0,1$.  The remaining
elementary coupling are  thus
$$\eqalignT{
&\A_1=(a^0,a^3)_0 \qquad &\A_2=(a^3,a^1)_0\qquad &\A_3=(a^1,a^0)_0 \cr
&\B_1=(a^0,a^2)_0\qquad &\B_2=(a^2a^2)_0\qquad &\B_3=(a^2,a^0)_0\cr
&\C_1=(a^0,a^1)_0\qquad &\C_2=(a^1,a^3)_0  \qquad&\C_3=(a^3,a^0)_0\cr
     &\D_1'=(a^2,a^1)_0\qquad &\D_2'=(a^1,a^1)_0\qquad &\D_3'=(a^1,a^2)_0 \cr
&\D_1=(a^2,a^3)_0\qquad &\D_2=(a^3,a^3)_0\qquad &\D_3=(a^3,a^2)_0 \cr
&\E_1=(a^0,a^1)_1\qquad &\E_2=(a^1,a^1)_1\qquad &\E_3=(a^1,a^0)_1\cr}\eq$$
(The explicit reexpression of the elementary couplings in terms of the LR
variables can
be found in [\BCM].)

The reconstruction of the polytope facets out of these vertices
reproduce the inequalities obtained in [\BCM]. These are
$$
\eqalignD{
k-\lambda_1-\lambda_2-\lambda_3-n_{33} &\geq 0 \qquad\qquad &
\lambda_3+n_{13}-n_{14}-n_{24} \geq 0 \cr
k-\lambda_1-\lambda_2-n_{11}-n_{14}+n_{34} &\geq 0 \qquad\qquad &
\lambda_3+n_{13}+n_{23}-n_{14}-n_{24}-n_{34} \geq 0 \cr
k-\lambda_1-n_{11}-n_{13}-n_{14} &\geq 0 \qquad\qquad &
n_{11}-n_{22} \geq 0 \cr
k-\lambda_1-\lambda_2-\lambda_3-n_{22}+n_{34} &\geq 0 \qquad\qquad &
n_{11}+n_{12}-n_{22}-n_{23} \geq 0 \cr
k-\lambda_1-\lambda_2-n_{14}-n_{22} &\geq 0 \qquad\qquad &
n_{11}+n_{12}+n_{13}-n_{22}-n_{23}-n_{24} \geq 0 \cr
k-\lambda_1-\lambda_2-\lambda_3-n_{11}+n_{24}+n_{34} &\geq 0 \qquad\qquad &
n_{22}-n_{33} \geq 0 \cr
k-\lambda_1-\lambda_2-\lambda_3+n_{12}-n_{22}-n_{23}+n_{34} &\geq 0
\qquad\qquad &
n_{22}+n_{23}-n_{33}-n_{34} \geq 0 \cr
k-\lambda_1-\lambda_2-n_{14}+n_{13}-n_{22}-n_{24} &\geq 0 \qquad\qquad &
n_{12} \geq 0 \cr
k-\lambda_1-\lambda_2-n_{11}-n_{14}+n_{23} &\geq 0 \qquad\qquad &
n_{13} \geq 0 \cr
2k-2\lambda_1-2\lambda_2-\lambda_3-n_{14}-n_{22}-n_{11}+n_{34} &\geq 0
\qquad\qquad &
n_{14} \geq 0 \cr
\lambda_1-n_{12} &\geq 0 \qquad\qquad &
n_{23} \geq 0 \cr
\lambda_2-n_{13} &\geq 0 \qquad\qquad &
n_{24} \geq 0 \cr
\lambda_2+n_{12}-n_{13}-n_{23} &\geq 0 \qquad\qquad &
n_{33} \geq 0 \cr
\lambda_3-n_{14} &\geq 0 \qquad\qquad &
n_{34} \geq 0 \cr}\eqlabel\qineq
$$

We then look for the symmetry  group of the resulting polytope
using the commutant of $Q$. The matrix $V'$, with the column ordering
$(\E_0, \,\A_1,
\cdots,\D_1',\cdots , \D_3, \,\E_1,\, \E_2,\,\E_3,\,\F)$, is
$$V'=k \pmatrix{
1 & 1 & 1 & 1 & 1 & 1 & 1 & 1 & 1 & 1 & 1 & 1 & 1 & 1 & 1 & 1 & 1 & 1 & 1 &
1 \cr
0 & 0 & 0 & 1 & 0 & 0 & 0 & 0 & 1 & 0 & 0 & 1 & 1 & 0 & 0 & 0 & {1 \over 2}
& 0 & 0 & {1 \over 2} \cr
0 & 0 & 0 & 0 & 0 & 1 & 1 & 0 & 0 & 0 & 1 & 0 & 0 & 1 & 0 & 0 & 0 & {1
\over 2} & {1 \over 2} & 0 \cr
0 & 0 & 1 & 0 & 0 & 0 & 0 & 0 & 0 & 1 & 0 & 0 & 0 & 0 & 1 & 1 & {1 \over 2}
& 0 & 0 & {1 \over 2} \cr
0 & 1 & 0 & 0 & 1 & 0 & 0 & 1 & 0 & 0 & 0 & 0 & 0 & 1 & 1 & 1 & 0 & {1
\over 2} & {1 \over 2} & {1 \over 2} \cr
0 & 0 & 0 & 0 & 0 & 0 & 0 & 0 & 1 & 0 & 0 & 1 & 1 & 0 & 0 & 0 & {1 \over 2}
& 0 & 0 & {1 \over 2} \cr
0 & 0 & 0 & 0 & 0 & 1 & 0 & 0 & 0 & 0 & 1 & 0 & 0 & 0 & 0 & 0 & 0 & 0 & {1
\over 2} & 0 \cr
0 & 0 & 1 & 0 & 0 & 0 & 0 & 0 & 0 & 0 & 0 & 0 & 0 & 0 & 0 & 0 & 0 & 0 & 0 &
0 \cr
0 & 1 & 0 & 0 & 1 & 0 & 0 & 0 & 0 & 0 & 0 & 0 & 0 & 0 & 1 & 0 & 0 & 0 & {1
\over 2} & 0 \cr
0 & 0 & 0 & 0 & 0 & 0 & 0 & 0 & 1 & 0 & 0 & 0 & 1 & 1 & 0 & 0 & 0 & {1
\over 2} & 0 & {1 \over 2} \cr
0 & 0 & 0 & 0 & 0 & 1 & 0 & 0 & 0 & 0 & 0 & 0 & 0 & 0 & 0 & 1 & {1 \over 2}
& 0 & 0 & 0 \cr
0 & 1 & 0 & 0 & 0 & 0 & 0 & 0 & 0 & 0 & 0 & 0 & 0 & 0 & 0 & 0 & 0 & 0 & 0 &
0 \cr
0 & 0 & 0 & 0 & 0 & 0 & 0 & 0 & 1 & 0 & 0 & 0 & 0 & 1 & 1 & 0 & 0 & 0 & {1
\over 2} & {1 \over 2} \cr
}\eq
$$
\font\sm cmr5
The matrix $Q'=  3240 \, Q$ reads
$${\sm
\pmatrix{ \hbox{\sm 2088} & \hbox{ \sm 0} & \hbox{ \sm 0} & \hbox{ \sm 0} &
\hbox{ \sm 168} & \hbox{ \sm 168} & \hbox{ \sm 168} &
\hbox{ \sm 600} &
\hbox{ \sm 600} & \hbox{ \sm 600} & \hbox{ \sm -360}
& \hbox{ \sm -360} &
\hbox{ \sm -360} & \hbox{ \sm -360} & \hbox{ \sm -
360} & \hbox{ \sm -360} &
\hbox{ \sm 384} & \hbox{ \sm 384} & \hbox{ \sm 384} &
\hbox{ \sm -144}\cr
\noalign{\medskip}\hbox{ \sm 0} & \hbox{ \sm 3240} &
\hbox{ \sm 0} & \hbox{
\sm 0
} & \hbox{ \sm 0} & \hbox{ \sm 0} & \hbox{ \sm 0} &
\hbox{ \sm 0} & \hbox{
\sm 0} & \hbox{ \sm 0} & \hbox{ \sm 0} & \hbox{ \sm
0} & \hbox{ \sm 0} &
\hbox{ \sm 0} & \hbox{ \sm 0} & \hbox{ \sm 0} &
\hbox{ \sm 0} & \hbox{ \sm
0} & \hbox{ \sm 0} & \hbox{ \sm 0}\cr
     \noalign{\medskip} \hbox{\sm 0} & \hbox{ \sm 0} &
\hbox{ \sm 3240} &
\hbox{ \sm 0} & \hbox{ \sm 0} & \hbox{ \sm 0} &
\hbox{ \sm 0} & \hbox{ \sm
0} & \hbox{ \sm 0} & \hbox{ \sm 0} & \hbox{ \sm 0} &
\hbox{ \sm 0} & \hbox{
\sm 0} & \hbox{ \sm 0} & \hbox{ \sm 0} & \hbox{ \sm
0} & \hbox{ \sm 0} &
\hbox{ \sm 0} & \hbox{ \sm 0} & \hbox{ \sm 0}\cr
     \noalign{\medskip}\hbox{\sm 0} & \hbox{ \sm 0} &
\hbox{ \sm 0} & \hbox{
\sm 3240} & \hbox{ \sm 0} & \hbox{ \sm 0} & \hbox{
\sm 0} & \hbox{ \sm 0} &
\hbox{ \sm 0} & \hbox{ \sm 0} & \hbox{ \sm 0} &
\hbox{ \sm 0} & \hbox{ \sm
0} & \hbox{ \sm 0} & \hbox{ \sm 0} & \hbox{ \sm 0} &
\hbox{ \sm 0} & \hbox{
\sm 0} & \hbox{ \sm 0
} & \hbox{ \sm 0}\cr \noalign{\medskip}\hbox{\sm 168}
& \hbox{ \sm 0} &
\hbox{ \sm 0} & \hbox{ \sm 0} & \hbox{ \sm 2338} &
\hbox{ \sm 43} & \hbox{
\sm 43} & \hbox{ \sm 250} & \hbox{ \sm -425} & \hbox{
\sm -425} & \hbox{
\sm -285} & \hbox{ \sm -15} & \hbox{ \sm 660} &
\hbox{ \sm -285} & \hbox{
\sm 660} & \hbox{ \sm -15} & \hbox{ \sm -56} & \hbox{
\sm 484} & \hbox{ \sm
484} & \hbox{ \sm -384}\cr
\noalign{\medskip}\hbox{\sm 168} & \hbox{ \sm 0}
& \hbox{ \sm 0} & \hbox{ \sm 0} & \hbox{ \sm 43} &
\hbox{ \sm 2338} &
\hbox{ \sm 43} & \hbox{ \sm -425} & \hbox{ \sm 250} &
\hbox{ \sm -425} &
\hbox{ \sm
660} & \hbox{ \sm -285} & \hbox{ \sm -15} & \hbox{
\sm -15} & \hbox{ \sm
-285} & \hbox{ \sm 660} & \hbox{ \sm 484} & \hbox{
\sm -56} & \hbox{ \sm
484} & \hbox{ \sm -384}\cr
     \noalign{\medskip}\hbox{ \sm 168} & \hbox{ \sm 0} &
\hbox{ \sm 0} & \hbox{
\sm 0} & \hbox{ \sm 43} & \hbox{ \sm 43} & \hbox{ \sm
2338} & \hbox{ \sm
-425} & \hbox{ \sm -425} & \hbox{ \sm 250} & \hbox{
\sm -15} & \hbox{ \sm
660} & \hbox{ \sm -285} & \hbox{ \sm 660} & \hbox{
\sm -15} & \hbox{ \sm
-285} & \hbox{ \sm 484} & \hbox{ \sm 484} & \hbox{
\sm -56} & \hbox{ \sm
-384}
\cr \noalign{\medskip}\hbox{\sm 600} & \hbox{ \sm 0}
& \hbox{ \sm 0} &
\hbox{ \sm 0} & \hbox{ \sm 250} & \hbox{ \sm -425} &
\hbox{ \sm -425} &
\hbox{ \sm 2050} & \hbox{ \sm -245} & \hbox{ \sm -
245} & \hbox{ \sm 255} &
\hbox{ \sm 525} & \hbox{ \sm -420} & \hbox{ \sm 255}
& \hbox{ \sm -420} &
\hbox{ \sm 525} & \hbox{ \sm -200} & \hbox{ \sm 340}
& \hbox{ \sm 340} &
\hbox{ \sm 480}\cr \noalign{\medskip}\hbox{\sm 600} &
\hbox{ \sm 0} &
\hbox{ \sm 0} & \hbox{ \sm 0} & \hbox{ \sm -425} &
\hbox{ \sm 250} & \hbox{
\sm -425} & \hbox{ \sm -245} & \hbox{ \sm 2050
} & \hbox{ \sm -245} & \hbox{ \sm -420} & \hbox{ \sm
255} & \hbox{ \sm 525}
& \hbox{ \sm 525} & \hbox{ \sm 255} & \hbox{ \sm -
420} & \hbox{ \sm 340} &
\hbox{ \sm -200} & \hbox{ \sm 340} & \hbox{ \sm
480}\cr
\noalign{\medskip}\hbox{\sm 600} & \hbox{ \sm 0} &
\hbox{ \sm 0} & \hbox{
\sm 0} & \hbox{ \sm -425} & \hbox{ \sm -425} & \hbox{
\sm 250} & \hbox{ \sm
-245} & \hbox{ \sm -245} & \hbox{ \sm 2050} & \hbox{
\sm 525} & \hbox{ \sm
-420} & \hbox{ \sm 255} & \hbox{ \sm -420} & \hbox{
\sm 525} & \hbox{ \sm
255} & \hbox{ \sm 340} & \hbox{ \sm 340} & \hbox{ \sm
-200} & \hbox{ \sm
480}
\cr \noalign{\medskip}\hbox{\sm -360} & \hbox{ \sm 0}
& \hbox{ \sm 0} &
\hbox{ \sm 0} & \hbox{ \sm -285} & \hbox{ \sm 660} &
\hbox{ \sm -15} &
\hbox{ \sm 255} & \hbox{ \sm -420} & \hbox{ \sm 525}
& \hbox{ \sm 2250} &
\hbox{ \sm 225} & \hbox{ \sm 225} & \hbox{ \sm -45} &
\hbox{ \sm -45} &
\hbox{ \sm -450} & \hbox{ \sm -420} & \hbox{ \sm 120}
& \hbox{ \sm 660} &
\hbox{ \sm 360}\cr \noalign{\medskip}\hbox{\sm -360}
& \hbox{ \sm 0} &
\hbox{ \sm 0} & \hbox{ \sm 0} & \hbox{ \sm -15} &
\hbox{ \sm -285} & \hbox{
\sm 660} & \hbox{ \sm 525} & \hbox{ \sm 255} & \hbox{
\sm -
420} & \hbox{ \sm 225} & \hbox{ \sm 2250} & \hbox{
\sm 225} & \hbox{ \sm
-450} & \hbox{ \sm -45} & \hbox{ \sm -45} & \hbox{
\sm 660} & \hbox{ \sm
-420} & \hbox{ \sm 120} & \hbox{ \sm 360}\cr
\noalign{\medskip}\hbox{\sm
-360} & \hbox{ \sm 0} & \hbox{ \sm 0} & \hbox{ \sm 0}
& \hbox{ \sm 660} &
\hbox{ \sm -15} & \hbox{ \sm -285} & \hbox{ \sm -420}
& \hbox{ \sm 525} &
\hbox{ \sm 255} & \hbox{ \sm 225} & \hbox{ \sm 225} &
\hbox{ \sm 2250} &
\hbox{ \sm -45} & \hbox{ \sm -450} & \hbox{ \sm -45}
& \hbox{ \sm 120} &
\hbox{ \sm 660} & \hbox{ \sm -420} & \hbox{ \sm 360}
\cr \noalign{\medskip}\hbox{\sm -360} & \hbox{ \sm 0}
& \hbox{ \sm 0} &
\hbox{ \sm 0} & \hbox{ \sm -285} & \hbox{ \sm -15} &
\hbox{ \sm 660} &
\hbox{ \sm 255} & \hbox{ \sm 525} & \hbox{ \sm -420}
& \hbox{ \sm -45} &
\hbox{ \sm -450} & \hbox{ \sm -45} & \hbox{ \sm 2250}
& \hbox{ \sm 225} &
\hbox{ \sm 225} & \hbox{ \sm -420} & \hbox{ \sm 660}
& \hbox{ \sm 120} &
\hbox{ \sm 360}\cr \noalign{\medskip}\hbox{\sm -360}
& \hbox{ \sm 0} &
\hbox{ \sm 0} & \hbox{ \sm 0} & \hbox{ \sm 660} &
\hbox{ \sm -285} & \hbox{
\sm -15} & \hbox{ \sm -420} & \hbox{ \sm 255} &
\hbox{ \sm
525} & \hbox{ \sm -45} & \hbox{ \sm -45} & \hbox{ \sm
-450} & \hbox{ \sm
225} & \hbox{ \sm 2250} & \hbox{ \sm 225} & \hbox{
\sm 120} & \hbox{ \sm
-420} & \hbox{ \sm 660} & \hbox{ \sm 360}\cr
\noalign{\medskip}\hbox{\sm
-360} & \hbox{ \sm 0} & \hbox{ \sm 0} & \hbox{ \sm 0}
& \hbox{ \sm -15} &
\hbox{ \sm 660} & \hbox{ \sm -285} & \hbox{ \sm 525}
& \hbox{ \sm -420} &
\hbox{ \sm 255} & \hbox{ \sm -450} & \hbox{ \sm -45}
& \hbox{ \sm -45} &
\hbox{ \sm 225} & \hbox{ \sm 225} & \hbox{ \sm 2250}
& \hbox{ \sm 660} &
\hbox{ \sm 120} & \hbox{ \sm -420} & \hbox{ \sm 360}
\cr \noalign{\medskip}\hbox{\sm 384} & \hbox{ \sm 0}
& \hbox{ \sm 0} &
\hbox{ \sm 0} & \hbox{ \sm -56} & \hbox{ \sm 484} &
\hbox{ \sm 484} &
\hbox{ \sm -200} & \hbox{ \sm 340} & \hbox{ \sm 340}
& \hbox{ \sm -420} &
\hbox{ \sm 660} & \hbox{ \sm 120} & \hbox{ \sm -420}
& \hbox{ \sm 120} &
\hbox{ \sm 660} & \hbox{ \sm 952} & \hbox{ \sm -128}
& \hbox{ \sm -128} &
\hbox{ \sm 48}\cr \noalign{\medskip}\hbox{\sm 384} &
\hbox{ \sm 0} & \hbox{
\sm 0} & \hbox{ \sm 0} & \hbox{ \sm 484} & \hbox{ \sm
-56} & \hbox{ \sm
484} & \hbox{ \sm 340} & \hbox{ \sm -200} & \hbox{
\sm 340} & \hbox{ \sm
120} & \hbox{ \sm -420} & \hbox{ \sm 660} & \hbox{
\sm 660} & \hbox{ \sm
-420} & \hbox{ \sm 120} & \hbox{ \sm -128} & \hbox{
\sm 952} & \hbox{ \sm
-128} & \hbox{ \sm 48}\cr
     \noalign{\medskip}\hbox{\sm 384} & \hbox{ \sm 0} &
\hbox{ \sm 0} & \hbox{
\sm 0} & \hbox{ \sm 484} & \hbox{ \sm 484} &
\hbox{\sm -56} & \hbox{ \sm
340} & \hbox{ \sm 340} & \hbox{ \sm -200} & \hbox{
\sm 660} & \hbox{ \sm
120} & \hbox{ \sm -420} & \hbox{ \sm 120} & \hbox{
\sm 660} & \hbox{ \sm
-420} & \hbox{ \sm -128} & \hbox{ \sm -128} & \hbox{
\sm 952} & \hbox{ \sm
48}
\cr \noalign{\medskip}\hbox{\sm -144} & \hbox{ \sm 0}
& \hbox{ \sm 0} &
\hbox{ \sm 0} & \hbox{ \sm -384} & \hbox{ \sm -384} &
\hbox{ \sm -384} &
\hbox{ \sm 480} & \hbox{ \sm 480} & \hbox{ \sm 480} &
\hbox{ \sm 360} &
\hbox{ \sm 360} & \hbox{ \sm 360} & \hbox{ \sm 360} &
\hbox{ \sm 360} &
\hbox{ \sm 360} & \hbox{ \sm 48} & \hbox{ \sm 48} &
\hbox{ \sm 48} & \hbox{
\sm 792 }  } }
\eq
$$
Again, considering the sets of entries in each row of
$Q$ we find that the
symmetry
group fixes the sets
$\{1\}$, $\{2, 3, 4\}$, $\{5, 6, 7\}$,
$\{8, 9, 10\}$, $\{11, 12, 13, 14, 15, 16\}$,
$\{17, 18, 19\}$ and
$\{20\}$.
We notice immediately that the permutations of
$2,\,3$ and 4 give
an $S_3$ symmetry that commutes
with all the other symmetries.

Let us suppose that $\sigma$ is a symmetry and let
$\tau$ be the induced
permutation
of
$\{11, 12, 13, 14, 15, 16\}$. The submatrix of $Q$
corresponding to rows
$\{11, 12, 13, 14, 15, 16\}$ and columns  $\{5, 6,
7\}$ is
$$U_1={1\over{3240}}
\pmatrix{ -285&660&-15\cr \noalign{\medskip}-15&-
285&660\cr
\noalign{\medskip}660&-15&-285\cr \noalign{\medskip}-
285&-15&660
\cr \noalign{\medskip}660&-285&-15\cr
\noalign{\medskip}-15&660&-285 }\eq
$$
while the submatrix for rows $\{11, 12, 13, 14, 15,
16\}$ and columns
$\{8, 9, 10\}$ is
$$U_2=
{1\over 3240}\pmatrix{  255&-420&525\cr
\noalign{\medskip}525&255&-420\cr
     \noalign{\medskip}-420&525&255\cr
\noalign{\medskip}255&525&-420
\cr \noalign{\medskip}-420&255&525\cr
\noalign{\medskip}525&-420&255 }
\eq
$$
and for rows $\{11, 12, 13, 14, 15, 16\}$ and columns
$\{17, 18, 19\}$ it is
$$U_3=
{1\over{3240}}
\pmatrix{ -420&120&660\cr \noalign{\medskip}660&-
420&120\cr
\noalign{\medskip}120&660&-420\cr \noalign{\medskip}-
420&660&120
\cr \noalign{\medskip}120&-420&660\cr
\noalign{\medskip}660&120&-420 }
\eq
$$
Since each of these matrices have rows with distinct
entries, we can
deduce the action
of $\sigma$ on each of the sets
$\{5, 6, 7\}$, $\{8, 9, 10\}$ and  $\{17, 18, 19\}$
from $\tau$.
Since $\sigma$ fixes 1 and 20
this determines $\sigma$ except for its action on
$2,\,3$ and 4, which, as noted
above, is arbitrary.

Thus, to find all symmetries, it suffices to find all
possible $\tau$. The
submatrix of $Q$ corresponding to rows $\{11, 12, 13,
14, 15, 16\}$ and
columns $\{11, 12, 13, 14, 15, 16\}$ is
$$
K=
{1\over{3240}}
\pmatrix{  2250&225&225&-45&-45&-450\cr
\noalign{\medskip}
225&2250&225&-450&-45&-45\cr
\noalign{\medskip}225&225&2250&-45&-450&-45
\cr \noalign{\medskip}-45&-450&-45&2250&225&225\cr
\noalign{\medskip}-45&-45&
-450&225&2250&225\cr \noalign{\medskip}-450&-45&-
45&225&225&2250 }\eq
$$
The permutation $\tau$ commutes with $K$ and $K$ has
an eigenvector
$w^\top=(-1, -1, -1, 1, 1, 1)$, the corresponding
eigenspace being
1-dimensional.
$\tau$ fixes this eigenspace and so either $\tau w =
w$ or $\tau w=-w$.
The group of all such permutations is $S_3\wr S_2$
with the
two $S_3$ groups permuting $\{11,12,13\}$ and
$\{14,15,16\}$ and the $S_2$
interchanging
them. Thus, every  $\sigma $ gives a $\tau$ in
$S_3\wr S_2$,
but it is not necessarily true that every element of
$S_3\wr S_2$
extends to a
symmetry of
$V'$. In fact, only a subgroup  can be
extended,
as we now show.

By trial and error, we can find two elements
$\alpha=(11,14)(12,16)(13,15)$ and
$\beta=(11,12,13)(14,15,16)$ of $S_3\wr S_2$ which
can be extended to the two symmetries
(6,7)(9,10)(11,14)(12,16)(13,15)(18,19) and
(5,6,7)(8,9,10)(11,12,13)(14,15,16)(17,18,19).
The group generated by $\alpha$ and $\beta$ turns
out to be isomorphic to $S_3$ (which does not
have the standard action).

Thus, we get a  group of symmetries isomorphic to
$S_3\times S_3$:
$$\eqalign{
G[\su(4)]= < (2,3),\, (2,3,4),\,&
(6,7)(9,10)(11,14)(12,16)(13,15)(18,19),\cr &
(5,6,7)(8,9,10)(11,12,13)(14,15,16)(17,18,19)>\cr}\eq
$$
There are no other symmetries for the following
reason. Suppose that $\sigma$ is a
symmetry and $\tau$ is the induced permutation on
$\{11, 12, 13, 14, 15, 16\}$.
Since the group generated by $\alpha$ and $\beta$ is
transitive on $\{11,
12, 13, 14,
15, 16\}$, there is some
$\tau'$ such that $\tau\tau'$ fixes 11. Then, by
considering the first row of
$U_1$, $\tau\tau'$ also fixes $\{5,6,7\}$. Hence
     $\tau\tau'$ must fix  $U_1$. But this implies
$\tau\tau'$ is the identity since $U_1$ has distinct
rows. Thus
$\tau^{-1}=\tau'$
and so $\sigma$ is in the group $G$.

With $x^{\top}=(k, \lambda_1, \lambda_2, \lambda_3,
n_{11}, n_{12}, n_{13},
     n_{14}, n_{22}, n_{23}, n_{24}, n_{33}, n_{34})$,
the changes of variables corresponding to the
generating symmetries are:
$$\eqalign{
&(2, 3):
x^{\top} \mapsto
(k, \lambda_1, \lambda_2, \lambda_3-n_{14}+n_{33},
n_{11}+n_{14}-n_{33},
     n_{12}, n_{13}, n_{33},\cr
&\qquad\qquad  n_{14}+n_{22}-n_{33}, n_{23}, n_{24},
n_{14}, n_{34})\cr
&(2, 3, 4):
x^{\top} \mapsto
    (k, n_{12}+n_{14}, \lambda_2, \lambda_3-
    n_{14}+n_{33},\lambda_1+
      n_{11}-n_{12}-n_{33}, n_{12}, n_{13},\cr
    &\qquad \qquad n_{33},\lambda_1-
    n_{12}+n_{22}-n_{33}, n_{23}, n_{24}, \lambda_1-n_{12},
    n_{34})\cr
&(11, 14)(12, 16)(13, 15)(6, 7)(18, 19)(9, 10):
x^{\top} \mapsto
(k, \lambda_1+\lambda_3-n_{12}-n_{14},\lambda_2,
n_{12}+n_{14},\cr
&\qquad \qquad
n_{11}+n_{12}+n_{13}-n_{24}-n_{34},
\lambda_3-n_{14}, \lambda_2-n_{13}, n_{14},
     n_{22}+n_{23}-n_{34}, \cr &\qquad
\qquad\lambda_3+n_{13}-n_{14}-n_{24},
\lambda_2+n_{12}-n_{13}-n_{23}, n_{33},
\lambda_3+n_{13}+n_{23}-n_{14}-n_{24}-n_{34})\hfill\cr
& (5, 6, 7)(8, 9, 10)(11, 12,
13)(14, 15, 16)(17, 18, 19): x^{\top}  \mapsto\cr
&\qquad\qquad
%
(k, \lambda_1+n_{11}+n_{13}-n_{22}-n_{23}-n_{24},
n_{22}+n_{23}+n_{24}-n_{33}-n_{34},\cr
&\qquad \qquad
n_{14}+n_{34},
\lambda_2+\lambda_3-n_{13}-n_{14}+n_{33},
n_{11}+n_{12}+n_{13}-n_{22}-n_{23}-n_{24},
\cr &\qquad\qquad
n_{22}+n_{23}-n_{33}-n_{34},
n_{14},
\lambda_2- n_{13}+n_{33},
n_{11}+n_{12}-n_{22}-n_{23},
n_{22}-n_{33},
n_{33}, n_{11}-n_{22})
\cr}\eq
$$

     Labelling the $\su(4)$ fusion inequalities of
     (\qineq) from 1 to 28 (row by row), the orbits on
the fusion basis
are:
$$\displaylines{
(2, 3): [[24, 27]]\hfill\cr
(2, 3, 4): [[11, 27, 24]]\hfill\cr
(11, 14)(12, 16)(13, 15)(6, 7)(18, 19)(9,
10):\hfill\cr\hfill
[[2, 9], [3, 6], [4, 8], [5, 7],  [12, 23], [13, 26],
[14, 22],
[15, 25], [16, 28], [17, 19], [20, 21]]\cr
(5, 6, 7)(8, 9, 10)(11, 12, 13)(14, 15, 16)(17, 18,
19):\hfill\cr\hfill
[[1, 6, 3], [2, 5, 4], [7, 9, 8], [12, 20, 26], [13,
21, 23],
[14, 17, 28], [15, 18, 25], [16, 19, 22]] \cr}
$$
$$\eq$$
where the orbits of length 1 have been omitted.

As already indicated,
severe constraints on the symmetries come from the
linear relations.
     And in fact
there are many such relations in the $\su(4)$ case.
The full list is
[\BCM]:
$$ \eqalignT{ &\E_0\D_j^{'}  \D_ k = \C_i \E_i \qquad
&\E_0\D_j \D_k^{'}
= \B_i\C_j
\C_k
\qquad &\E_i \E_j  = \E_0\B_k \D_k \D_k^{'} \cr
&\D_i \E_i  = \C_j \B_k \D_k \qquad
&\D_i^{'}\E_i   = \B_j \D_j^{'} \C_k \qquad &\E_0\F=
\C_1\C_2\C_3 {} \cr}
\eqlabel\mmmhb$$
with $i,j,k$ a cyclic permutation of $1,2,3$. The
large number of
relations, and more precisely, the fact that they mix
elementary couplings with
different threshold levels, is responsible for the
absence of symmetries involving the level.
All the
symmetries found above leave this set of relations
invariant. For instance, the
$S_3$ group generated by (2,3) and (2,3,4) is the
permutation group of the three
$\A_i$'s, which do not appear in the relations.

With the vector $y^{\top}$ defined as
$y^{\top}=(k,\lambda_1,\lambda_2,\lambda_3,\mu_1,\mu_2
,\mu_3, \nu_1,\nu_2,\nu_3,n_{12}, n_{14}, n_{33})
$, we can rewrite the symmetries of the fusion basis
as:
$$\eqalignD{& (2,3) :  y^{\top} \rightarrow
(k, \lambda_1, \lambda_2, \lambda_3-n_{14}+n_{33},
\mu_1-n_{14}+n_{33},
\mu_2,\mu_3+n_{14}-n_{33}, \nu_1+n_{14}-n_{33}, \cr
     & \qquad\qquad \nu_2, \nu_3, n_{12}, n_{14}, n_{33}
)\cr
& (2,3,4) : y^{\top}
\nu_1+\lambda_1-n_{12}-n_{33},\nu_2,\nu_3-\lambda_1+n_{12}+n_{14},
\rightarrow (k, n_{12}+n_{33},\lambda_2,\lambda_1+\lambda_3-n_{12}-n_{14},
\lambda_1+\mu_1- n_{12}-n_{14},
\mu_2,
\mu_3+n_{14}-n_{33}, \cr
& \qquad\qquad
\nu_1+n_{14}-n_{33},\nu_2,-\lambda_1+\nu_3+n_{12}+n_{33},
n_{12},\lambda_1-n_{12},n_{14})\cr
& (11,14) (12,16) (13,15) (6,7) (18,19) (9,10) :
y^{\top} \rightarrow
(k,\lambda_1+\lambda_3-n_{12}-n_{14},\lambda_2, n_{12}+n_{14},\cr
& \qquad\qquad \nu_3-\lambda_1+n_{12}+n_{14},\nu_2, \nu_1,
\mu_3,\mu_2,\lambda_1+\mu_1-n_{12}-
n_{14},\lambda_3-n_{14},n_{14},n_{33})\cr
& (5,6,7) (8,9,10) (11,12,13) (14,15,16) (17,18,19) :
y^{\top} \rightarrow
%
(k, \lambda_1+\nu_1-n_{12}-n_{33},
\nu_2,\cr & \qquad\qquad
-\lambda_1+\nu_3+n_{12}+n_{14},
n_{12}+n_{14},\lambda_2,
\lambda_3-n_{14}+n_{33},
\mu_1-n_{14}+n_{33},
\mu_2, \cr
& \qquad\qquad
\lambda_1+\mu_3-n_{12}-n_{33},
\nu_1-n_{33},n_{14},n_{33})\cr
}\eq $$

\noindent These symmetries do not correspond to a fusion-coefficient
symmetries.

\newsec{Conclusion}

In this paper we presented the concept of fusion bases,
first  introduced in
[\BCM],
from a novel point of view as the set of facets of a
polytope.
This reformulation gives
access to the powerful computer programs that have
been developed for
generating
     facets out of the vertices.  Moreover, by
reformulating the problem in a
geometrical way, we were led to the study of the
affine symmetry group of the fusion
polytope, introduced here for the first time.
We developed simple tools for studying
this group analytically for the lowest rank affine
Lie algebras.
We also defined the vertex symmetry group
of a polytope and noted that for fusion polytopes the
affine and vertex symmetry groups appear to be
the same -- a property which does not hold for
general polytopes.

The order of the vertex symmetry group of the fusion polytope of the
     lowest rank affine Lie algebras was found to be  24 for
$\su(2)$, 432 for
$\su(3)$,  36
for $\su(4)$, and 4 for $\sp(4)$. Comparing $\su(2)$ and $\su(3)$,
it is natural to see an increase of
the order of the group with the rank since the number of
vertices increases rapidly. However, it might be
surprising to observe this drastic reduction in the order when passing from
$\su(3)$ to $\su(4)$. The reason is
that the number of linear relations, which have to be preserved by the
symmetry transformations, also increases rapidly with the rank.

We should stress that a fusion polytope is rather
special type of polytope
in that
its vertices are the elementary solutions of the
facets, which is far from
being a
generic property of polytopes.

It is an interesting open problem to try to generate
the full set of fusion
inequalities, or equivalently, to give a generic
description of the fusion
polytope,
from a general Lie algebraic point of view.

On the other hand, we could ask whether the polytope we have obtained,
whose description relies heavily on the LR variables,  is the `genuine'
fusion polytope or whether it is just one among a variety of polytopes.
In that vein, we  note that Rasmussen and Walton [\ref{J. Rasmussen and
M.A. Walton, {\it Affine su(3) and su(4) fusion multiplicities as polytope
volumes}, arXiv:hep-th/0106287. (2001)}\refname\RW]
have recently also developed a polytope interpretation
of $\su(3)$ and $\su(4)$ fusion coefficients using a different
approach from ours. Prompted by the referee we
have investigated the relationship with the polytopes found in this
paper. We find, perhaps surprisingly, after removing redundant variables
and making a suitable change of coordinates that the two sets of polytopes
coincide. This suggests that the polytopes might be in
some sense unique.

Finally, concerning the symmetry analysis, we stress that we have
restricted our analysis to a special class of symmetries, namely those
which exist at all level. Whether symmetries at particular levels or
even  symmetries that relate fusion polytopes at different levels can be
unraveled by our method remains to be studied.

\vskip0.3cm
\centerline{\bf Acknowledgment}

Part of this work was done during the workshop {\it Quantum
integrability} 2001 at the CRM, Universit\'e de Montr\'eal. We also
acknowledge the financial support of  NSERC (Canada) and FCAR (Qu\'ebec).

\vfill\eject
\centerline{\bf REFERENCES}
\vskip 1cm
\immediate\closeout\refs \vskip 0.5cm
      \message{References}\input references
\vfill\eject

\end